\documentclass[journal,usenames,dvipsnames]{IEEEtran}

\usepackage[english]{babel}
\usepackage{blindtext}

\usepackage{multirow}
\usepackage{makecell}

\usepackage{cite}
\usepackage{amsmath,amsfonts}
\usepackage{mathtools}
\usepackage{algorithmic}
\usepackage{graphicx}
\usepackage{tikz}
\usetikzlibrary{er,positioning,decorations,decorations.pathreplacing,decorations.pathmorphing,shapes,fadings,fit,backgrounds,shapes,calc,arrows,arrows.meta,patterns}
\definecolor{csgreen}{HTML}{bfe7dc}
\definecolor{csGREEN}{HTML}{009e73}
\definecolor{csblue}{HTML}{d5edfa}
\definecolor{csBLUE}{HTML}{56b4e9}
\definecolor{csZSK}{HTML}{0000ff}
\definecolor{csKSK}{HTML}{ffa500}
\tikzset{
	font=\footnotesize,
	>=Stealth,
	pics/CS/.style args={#1/#2/#3/#4/#5}{
		code = {%
			\draw[thick,#5] (0,0) -- ++(#1,0);
			\coordinate (sig start) at (#2,0);
			\coordinate (sig end) at ($(sig start)+(#3,0)$);
			\coordinate (sig turn) at ($(sig start)!#4!(sig end)$);
			\begin{scope}
				\clip[postaction={fill=csblue,draw=csBLUE,thick}] ($(sig turn)+(0,-.5em)$) rectangle ($(sig end)+(0,.5em)$);
			\end{scope}
			\begin{scope}
				\clip[postaction={fill=csgreen,draw=csGREEN,thick}] ($(sig start)+(0,-.5em)$) rectangle ($(sig turn)+(0,.5em)$);
			\end{scope}
			\path[tips,{Bar[width=1em]}-{Bar[width=1em]},thick,#5] (0,0) -- ++(#1,0);
		}
	},
	pics/ZSK/.style args={#1/#2/#3/#4}{
		code = {%
			\pic {CS=#1/#2/#3/#4/csZSK};
		}
	},
	pics/KSK/.style args={#1/#2/#3/#4/#5/#6}{
		code = {%
			\pic {CS=#1/#4/#5/#6/csKSK};
			\draw[thick,orange,arrows={|[harpoon]-|[harpoon,swap]}] (#2,1.1em) -- ++(#3,0);
		}
	}
}

\usetikzlibrary{external}
\usepackage{textcomp}
\usepackage{colortbl}
\usepackage[hyphens]{url}
\usepackage[labelformat=simple]{subcaption}

\usepackage{dblfloatfix} %



\usepackage{pgfplots}
\usepackage{pgfplotstable}
\usepgfplotslibrary{colorbrewer}

\usepackage{tabularx,booktabs}
\usepackage{arydshln}
\newcommand*\dhline{\specialrule{0pt}{1pt}{0pt}\hdashline[.4pt/3pt]\specialrule{0pt}{0pt}{1pt}}
\newcommand{\dcline}[1]{\specialrule{0pt}{1pt}{0pt}\cdashline{#1}[.4pt/3pt]\specialrule{0pt}{0pt}{1pt}}

\usepackage[hidelinks]{hyperref}
\usepackage{amsmath}
\usepackage{amssymb}
\usepackage{balance}
\usepackage{comment}
\usepackage{xcolor}

\usepackage[absolute,showboxes]{textpos}

\let\orgautoref\autoref
\renewcommand{\autoref}
{\def\sectionautorefname{Section}%
\def\subsectionautorefname{Section}%
\def\subsubsectionautorefname{Section}%
\orgautoref}

\usepackage{pifont}

\usepackage{wasysym}

\usepackage{xspace}
\newcommand{\etal}{\textit{et al.}}
\newcommand{\eg}{\textit{e.g.,}~}
\newcommand{\ie}{\textit{i.e.,}~}

\newcommand{\one}{({\em i})\xspace}
\newcommand{\two}{({\em ii})\xspace}
\newcommand{\three}{({\em iii})\xspace}

\makeatletter
\renewcommand{\paragraph}[1]{\vspace*{0.03in}\noindent{\bf #1.}\hspace{0.25ex \@plus1ex \@minus.2ex}}
\makeatother

\AtBeginDocument{%
  \providecommand\BibTeX{{%
    \normalfont B\kern-0.5em{\scshape i\kern-0.25em b}\kern-0.8em\TeX}}}

\begin{document}

\title{From the Beginning: Key Transitions in the\\ First 15 Years of DNSSEC}

\author{Eric Osterweil,  Pouyan Fotouhi Tehrani, Thomas~C.~Schmidt, and~Matthias~W\"ahlisch%
        \thanks{Eric Osterweil is with George Mason University, USA (e-mail: eoster@gmu.edu).}%
        \thanks{Pouyan Fotouhi Tehrani is with Weizenbaum Institute and Fraunhofer FOKUS, Germany (e-mail: pft@ieee.org).}%
        \thanks{Thomas C. Schmidt is with the Hamburg University of Applied Sciences~(HAW), Germany (e-mail: t.schmidt@haw-hamburg.de).}%
        \thanks{Matthias~W\"ahlisch is with Freie Universit\"at Berlin, Germany. (e-mail: m.waehlisch@fu-berlin.de).}%
      }

\maketitle

\definecolor{boxgray}{rgb}{0.93,0.93,0.93}
 \textblockcolor{boxgray}
 \setlength{\TPboxrulesize}{0.7pt}
 \setlength{\TPHorizModule}{\paperwidth}
 \setlength{\TPVertModule}{\paperheight}
 \TPMargin{5pt}
 \begin{textblock}{0.8}(0.1,0.01)
   \noindent
   \footnotesize
   If you refer to this paper, please cite the peer-reviewed publication: E. Osterweil, P. Fotouhi Tehrani, T. C. Schmidt and M. W\"ahlisch, ``From the Beginning: Key Transitions in the First 15 Years of DNSSEC,'' in \emph{IEEE Transactions on Network and Service Management}, vol. 19, no. 4, pp. 5265--5283, Dec. 2022, doi: \href{https://doi.org/10.1109/TNSM.2022.3195406}{10.1109/TNSM.2022.3195406}
\end{textblock}

\begin{abstract}

  When the global rollout of the DNS Security Extensions (DNSSEC) began in 2005, a first-of-its-kind trial started:
  The complexity of a core Internet protocol was magnified in favor of better security for the overall Internet. 
Thereby, the scale of the loosely-federated delegation in DNS became an unprecedented cryptographic key management challenge.  
  Though fundamental for current and future operational success, our community lacks a clear notion of how to empirically evaluate the process of securely transitioning keys.

  In this paper, we propose two building blocks to formally characterize and assess key transitions.
	First, the \emph{anatomy of key transitions}, \ie measurable and well-defined properties of key changes; and second, a novel \emph{classification model} based on this anatomy for describing key transition practices in abstract terms.
  This abstraction allows for classifying operational behavior.
  We apply our proposed transition anatomy and transition classes to describe the global DNSSEC deployment.
  Specifically, we use measurements from the first 15 years of the DNSSEC rollout to 
  detect and understand which key transitions have been used to what degree and which rates of errors and warnings occurred.
  In contrast to prior work, we consider all possible transitions and not only 1:1 key rollovers.
  Our results show measurable gaps between prescribed key management processes and key transitions in the wild.
  We also find evidence that such noncompliant transitions are needed in operations.

\end{abstract}

\begin{IEEEkeywords}
Domain Name System, DNSSEC, PKI, key rollover, Internet measurement, Information security.
\end{IEEEkeywords}

\section{Introduction}

Key transitions are critical for cryptographically enhanced infrastructures at the Internet-scale.
The Internet  is composed of loosely-federated administrative domains, and managing cryptographic keys under those conditions raises operational challenges in particular whenever domains depend on one another.  Mismanagement of cryptography can lead to security shortfalls in those systems and infrastructures that depend on them~(\eg \cite{l-edcs-13,k-ddd-13,p-cpsci-16,t-llir-20}).
Regarding DNS, Paul Mockapetris and Kevin J. Dunlap (1988) wrote, ``Distributing authority for a database does not distribute a corresponding amount of expertise''~\cite{mockapetris-sigcomm88}. 
DNSSEC, which began its global deployment in 2005, implicitly raised the question: \emph{Can} distributing responsibility to manage cryptographic material \emph{teach us} the corresponding amount of expertise?

In DNSSEC it is common to periodically change the cryptographic keys in use.
In some cases this is done as a hygienic prescription, in others it can be an emergency response to a security event (such as  a key compromise or a cryptosystem weakness).
The changing of keys is a process called key transitioning (or key rollover).
In this process, a system gracefully transitions from using keys that are \emph{departing}, \ie being removed, to using keys that are \emph{remaining}, \ie kept unchanged or being newly added, while ensuring continuity of protections during the change.
Following structured and validated process models for key transitions is critical for maintaining the security assurances of the overall system.
Different infrastructures dictate different processes and prescribe them in different ways~\cite{rfc5011,rfc7583,rfc6480,barker2007nist}.
While guidance for DNSSEC key life cycle management and timing exists in RFCs, a more foundational evaluation framework is missing so that the community can
quantify and evaluate \emph{all} operational aspects.
Such a framework would not only allow us to compare 
real deployments to prescribed guidance but to each others as well.

\begin{figure}
	\centering
	\newcommand{\timepoint}{circle (1.5pt)}

\tikzexternaldisable
\begin{tikzpicture}[
	>=Stealth,
	font=\tiny,
	good/.style={MidnightBlue},
	bad/.style={BrickRed},
	thiswork/.style={BurntOrange}
	]
\draw[|->] (-10pt,0) coordinate (start) -- ++(230pt,0) node[above,xshift=-2pt] {$year$};
\foreach \x in {0,14,28,...,210}{
	\draw[fill,gray] (\x pt,0) circle (1pt);
}

\draw[fill] (0,0) node[below] {'05} \timepoint;
\draw[->,good] (0,0) -- ++(0,.5) node[above,xshift=8pt] {\shortstack{DNSSEC\\Deployment\\+ \texttt{.se} signed}};

\draw[fill] (42pt,0) node[below] {'08} \timepoint;
\draw[->,bad] (42pt,0) -- ++(0,.5) node[above] {\shortstack{Kaminsky\\Attack}};

\draw[fill] (70pt,0) node[below] {'10} \timepoint;
\draw[->,good] (70pt,0) -- ++(0,1.15) node[above] {\shortstack{Root + \texttt{.net} signed\\DURZ rollover}};

\draw[fill] (84pt,0) node[below] {'11} \timepoint;
\draw[->,good] (84pt,0) -- ++(0,.5) node[above] {\shortstack{\texttt{.com}\\signed}};

\draw[fill] (168pt,0) node[below] {'17} \timepoint;
\draw[fill] (196pt,0) node[below] {'19} \timepoint;
\draw[|-|,bad] (168pt,1.5) -- node[above] {Sea Turtle} ++(28pt,0);

\draw[fill] (182pt,0) node[below] {'18} \timepoint;
\draw[->,good] (182pt,0) -- ++(0,.5) node[above] {\shortstack{Root KSK rollover}};

\draw[->,bad] (182pt,.85) -- ++(0,.25) node[above,yshift=-3pt] {\shortstack{DNSpionage}};

\draw[fill] (210pt,0) node[below] {'20} \timepoint;
\draw[->,bad] (210pt,0) -- ++(0,.75) node[above,] {SAD DNS};

\draw[|-|,thick,thiswork] (0,-1.75) -- node[midway,fill=white] {\bfseries{This work coverage}} ++(210pt,0);

\draw[|-|] (0,-.5) -- node[below] {Osterweil~\etal~\cite{Osterweil:2008:QOS:1452520.1452548}} ++(42pt,0);

\draw[|-|] ($(70pt,-.5)+(-2pt,0)$) -- ++(4pt,0) node[below] {Deccio~\etal~\cite{Deccio_2012}};

\draw[fill] (112pt,0) node[below] {'13} \timepoint;
\draw[|-|] ($(112pt,-.5)+(-2pt,0)$) -- ++(4pt,0) node[below] {Lian~\etal~\cite{182950}};

\draw[fill] (140pt,0) node[below] {'15} \timepoint;
\draw[|-|] (140pt,-1.25) node[left] {Chung~\etal~\cite{chung2017longitudinal}} -- ++(28pt,0);

\draw[|-|] ($(168pt,-.5)+(-2pt,0)$) -- ++(4pt,0) node[below] {M{\"u}ller~\etal~\cite{muller2019rolling}};

\draw[|-|] (168pt, -1) node[left] {van Rijswijk-Deij~\etal~\cite{10.1145/3123878.3131987}} -- ++(14pt,0);

\draw[|-|] (168pt, -1.5) node[left] (top rel) {M{\"u}ller~\etal~\cite{Muller:2019:RRR:3355369.3355570}} -- ++(28pt,0);

;
\draw[decorate,decoration=brace] (-8pt,-1.85) -- node[sloped,above] {\sffamily\bfseries{Data coverage}} ++(0,1.75);

\draw[decorate,decoration=brace] (-8pt,.1) -- node[sloped,above] {\sffamily\bfseries{DNS(SEC) Events}} ++(0,1.75);
\end{tikzpicture}
\tikzexternalenable
	\caption{Notable DNS(SEC) deployment events (blue) and security incidents~(red) during the measurement periods of related work (black) and this work (orange).}
	\label{fig:timeline}
\end{figure}
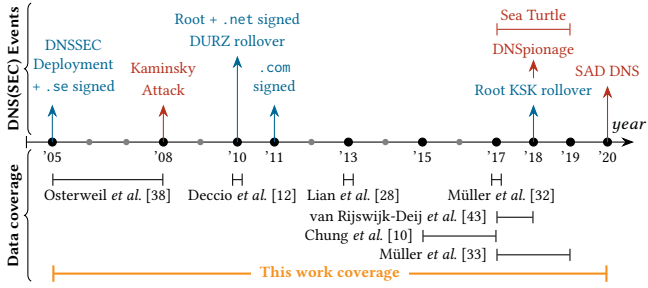

In this work, we propose a novel method to precisely model DNSSEC key transitions and apply this to analyze and classify the data from the first 15 years of DNSSEC deployment (2005--2020).
We define the composing elements required by our model as the \emph{anatomy of DNSSEC key transitions}. 
Furthermore, we propose a measurement methodology to quantify key transitions observed in the wild. 
By using our anatomy and transition model, we are able to model key transition behaviors in the wild
    from both RFCs~\cite{rfc5011,rfc7583} and from related work in the literature~\cite{wang2014emergency}.
    Our measurements cover $\approx$~19~million key transitions.
    They reveal a significant amount of operational heterogeneity, many of which deviate significantly from standards without necessarily degrading security.  
Our contributions in this work are threefold:

\begin{enumerate}
  \item {\bf Anatomy:} 
    We  examine the timing features of keys while transitioning and 
    propose an \emph{anatomy of DNSSEC key transitions},
    which defines a candidate set of measures that are necessary to measurably characterize key transitions.
  \item {\bf Transition classification:}  Based on our proposed anatomy, we present a novel methodology, which we use to concisely quantify and analyze key transitions.
  \item {\bf Longitudinal study:} To illustrate the generality of this work, we present measurements of operational key transitions that span 15 years 
    of DNSSEC deployment, covering a number of notable events that 
    have not been fully analyzed by
    related works  
    (see \autoref{fig:timeline}).
\end{enumerate}

The remainder of this paper is organized as follows. Section~\ref{sec:bg} summarizes background about DNSSEC.
In \autoref{sec:key-trans}, we propose our key transitions anatomy and explain how we construct our model of transition.
We follow this by proposing a security analysis of DNSSEC key life cycle and key transitions in Section~\ref{sec:sec-analysis}.
We present the measurement corpus used in this work in Section~\ref{sec:corpus}.
Section~\ref{sec:methodology} details our methodology to derive a continuous model of the  DNSSEC key lifetimes  from discrete observations.
Section~\ref{sec:classification} introduces our approach to classify
life cycle management and transitions of keys.
Based on that, we proceed to our extensive use case study in Section~\ref{sec:meas}.
We cover related work in Section~\ref{sec:relwork} and discuss our results in Section~\ref{sec:disc-disc}.
Finally, we conclude in Section~\ref{sec:disc}.

\section{Background}\label{sec:bg}

The DNS is a hierarchically administered global database of Resource Records (RRs), which are inserted and removed whenever operators choose.
The design of the DNS allows the administrator of any sub-tree (called a DNS \emph{zone}) to delegate the management authority of any branch under their zone to another authority.
Delegations are implemented when a zone parent adds Name Server entries, \ie \texttt{NS} RRs, that point to the DNS name servers of that sub-zone.
This hierarchical delegation allows administrators to operate their zones independently, and requires only a one-time coordination as long as the name servers remain the same.

To compensate for a number of security threats~(see \cite{bellovin1995using, RFC3833}), DNS evolved to have the \textit{DNS Security Extensions} (DNSSEC), whose specifications underwent their final round of standardization~\cite{rfc4033,rfc4034,rfc4035}
in 2005.
DNSSEC overloads the hierarchical namespace of DNS with cryptographic key learning and verification.
By design, DNSSEC-enabled zones generate and manage their own cryptographic key pairs using any set of DNSSEC
standardized cryptosystems.
Operators then encode the public portion of their key pair in a new RR type, {\tt DNSKEY}.
A DNSSEC signature, an \texttt{RRSIG}, is generated by the respective private portion of a \texttt{DNSKEY} over each set of same-type RRs, called an \emph{RRSet}, and is always returned with each DNSSEC~response.
An \texttt{RRSIG}  specifies its \emph{inception} and \emph{expiration} times to limit its period of validity and to resist replay attacks.
As these dates are defined in absolute values, DNSSEC implicitly requires ``loose time synchronization'' between authoritative nameservers and validating resolvers~\cite{RFC3833}.

DNSSEC specifies that zones should manage two separate classes of {\tt DNSKEY}s: Zone Signing Keys (ZSKs) and Key Signing Keys (KSKs).
While the cryptographic material used for these keys is fundamentally identical, their key life cycle management and roles are distinct: whereas ZSKs are used to sign all of a zone's contents (\eg {\tt A} and {\tt NS} records), KSKs are only used to sign {\tt DNSKEY} RRSets.
The root zone uses a well-known, self-signed KSK. All other zones need to have their KSKs authorized by a Delegation Signer ({\tt DS}) record in their parent zone so that keys at each zone can be globally verified by Relying Party~(RP) software (also called validating recursive resolvers, or just validators).
In this way, validators use the KSK of the DNS Root zone as a Trust Anchor (TA) and point of departure to construct unambiguous verifiable paths to any DNSSEC zone in the hierarchy through recursive tracing of secure delegations.
This follows the same way DNS zones are normally resolved recursively via {\tt NS} records.
The entire secure delegation chain from the DNS root to delegated zones, the \emph{chain of trust}, is the element that creates operational dependence between the cryptographic management of zones and their parents.

To validate a single RRSet, an RP relies on various pieces of information (\ie \texttt{DNSKEY}s, \texttt{RRSIG}s, \texttt{DS} records) from possibly different zones and sources (\ie authoritative servers or caches).
Validation of an RRSet in a given zone succeeds if \one a verifiable path to that zone exists and \two the signature over that RRSet is valid. 
Breaking the first requirement causes an RP to determine the state of data as insecure, while neglecting the second ends with a bogus state~\cite{rfc4033}.
These two requirements also constrain when and how operators can insert and remove RRs without defeating  the protection provided by DNSSEC.
Note that a valid signature presupposes that formal cryptographic requirements are met (\eg digests are correctly calculated), the key generating the signature is included in the \texttt{DNSKEY} RRSet, and the set is available and valid for at least the total validity period of the signature.
We provide a thorough discussion on temporal constraints with a focus on key transitions in \autoref{sec:sec-analysis}. 

\section{Modeling Key Transitions}\label{sec:key-trans}
\begin{figure}
	\centering
	\tiny
	\includegraphics{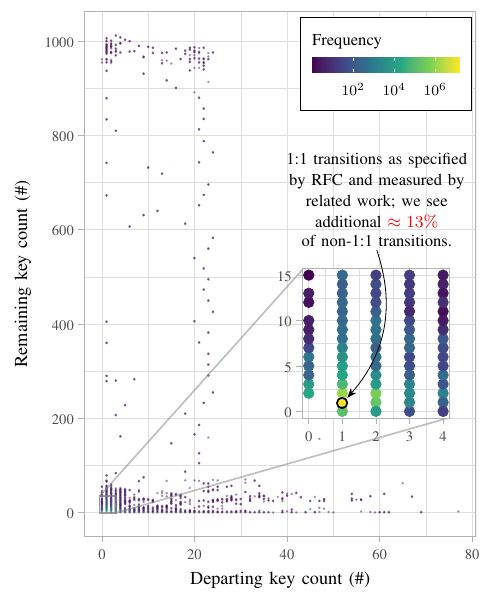}
	\caption{Frequencies of key transitions with different cardinalities between 2005--2020.}
	\label{fig:tot-card-chng}
\end{figure}

Key transition refers to the procedure of modifying the set of valid DNS keys over time.
The growing literature discusses DNSSEC key transitions in terms of key ``rollovers''~\cite{muller2019rolling,Muller:2019:RRR:3355369.3355570}, \ie a single new key replacing the only existing key.
Based on a consistent monitoring of the first 15 DNSSEC years, we find evidence that the global DNSSEC deployment follows a different reality as is depicted in \autoref{fig:tot-card-chng}. Besides the expected one-to-one transitions, a notable portion (13\%)  of transitions involve more than two keys and require a more expressive model.
To reflect this reality, we introduce a generic model of key transitions.

In our generic model, a transition is characterized by an effective change of the DNSSEC key set, \ie 
a transition succeeds only if the removal of one or more \emph{departing keys} results in an altered set of \emph{remaining keys}.
It should be noted that the set of remaining keys includes both newly added keys (if any) and  keys which existed throughout the transition.
Our model allows us to evaluate both simple key rollovers involving a single departing and a single new key as well as more complex key transitions involving multiple keys, also seen in the wild.

In the following, we introduce a temporal model of DNS keys, which we 
use to define a
\emph{transition anatomy} and provide a method to capture the semantics of key transitions in terms of \emph{transition classes}.

\subsection{Anatomy of a Key Transition}\label{sec:key-trans-anat}

Key transitions are measurable
through changes in \texttt{DNSKEY} resource records, and their respective \texttt{RRSIG} records as published by the authoritative servers of the records.
In the case of KSKs,  changes in \texttt{DS} records require monitoring in the parent zone.
While such changes are present, we consider a transition as ongoing.

\begin{figure}%
	\centering
	\includegraphics{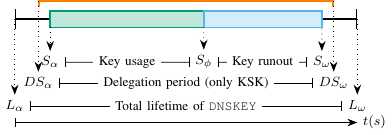}
	\caption{A temporal model spanning the \textit{total lifetime} of a \texttt{DNSKEY}. \texttt{RRSIG}s generated by the key define active \textit{key usage}, during \textit{key runout} no new signatures are created while existing ones remain valid,  and \texttt{RRSIG}(s) covering respective \texttt{DS} records define the key \textit{delegation period}.}
	\label{fig:temporal-model-dnskey}
\end{figure}

Before we describe the temporal aspects of \texttt{DNSKEY} transitions, we first need to define a life cycle model for \texttt{DNSKEY}s, which adequately describes a key existence from its inception throughout its usage, run-out, and its expiration. 
These four phases are depicted in \autoref{fig:temporal-model-dnskey}.

The only temporal information that is explicitly expressed in DNSSEC about resource records is encoded in \one \texttt{TTL} values and \two the validity period as defined by \texttt{RRSIG}s.
\texttt{TTL} values are indicators used by caching resolvers to \emph{locally} determine a time window after which a record should be considered stale and flushed from the cache.
The main purpose of TTLs is to establish and maintain eventual consistency in caches~\cite[\S8.1]{rfc4033}.
In contrast, the validity of \texttt{DNSKEY}s as denoted through \texttt{RRSIG}s provides information that can be used to reconstruct the life cycle of any key.
Accordingly, we create a life cycle model of keys using \one signatures \emph{over} and \two signatures \emph{generated by} those keys.
Using \texttt{RRSIG}s \emph{over} a key (recall that a single \texttt{DNSKEY} can be signed multiple times throughout its lifetime) the total lifetime of that key can be measured.
We denote the earliest and latest point in time when the key was singed as $L_\alpha$ and $L_\omega$, respectively.
The \texttt{RRSIG}s \emph{generated} by a \texttt{DNSKEY}, in turn, can be used to determine when the key was active and in use.
Formally we denote the interval from earliest and latest times that data was verifiable by this key~$[S_\alpha,S_\omega]$, the time when the key stopped generating new signatures~$S_\phi$, and subsequently the duration in which the key was only used to verify existing signatures~$[S_\phi,S_\omega]$.
Additionally, for KSKs only, the signatures over \texttt{DS} records of the parent zone are used to infer the period in which the parent zone was securely delegating to this key ($[DS_\alpha,DS_\omega]$).

\begin{figure}
	\tiny
	\includegraphics{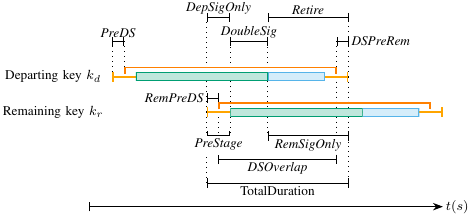}
	\caption{Anatomy of a $1:1$ key transition.}
	\label{fig:trans-anatomy}
\end{figure}

We now use this temporal model to characterize a simple 1:1 key rollover.
To this end we define a \emph{transition anatomy} based on the following ten~features, which 
 include measures regarding a remaining key $k_r$ and a departing key $k_d$, as visualized in \autoref{fig:trans-anatomy}:

\begin{enumerate}
	\item \textit{PreDS}: When a corresponding \texttt{DS} record was introduced for $k_d$ in the parent zone.
	\item \textit{DoubleSig}: The period during the transition when both departing and remaining keys were \emph{actively} signing.
	\item \textit{PreStage}: The time during which the remaining key was valid, but \emph{before} being used to verify zone data.
	\item \textit{DepSigOnly}: The duration during the key transition when \emph{only} the departing key was in use.
	\item \textit{Retire}: The duration during the key transition when signatures generated by departing key run out (run out described in Figure~\ref{fig:temporal-model-dnskey}).
	\item \textit{DSOverlap}: The duration (if at all) that {\tt DS}(es) for the departing and remaining keys overlapped.
	\item \textit{RemSigOnly}: The duration during the key transition when \emph{only} the remaining key was in use.
	\item \textit{DsPreRem}: If the departing key was covered by a {\tt DS}, the amount of time that the key was valid \emph{after} {\tt DS}(es) no longer delegated to it.
	\item \textit{RemPreDS}: When a corresponding \texttt{DS} record was introduced for the remaining key $k_r$ in the parent zone.
	\item \textit{TotalDuration}: Duration of the entire transition.
\end{enumerate}

\begin{table}
	\caption{Temporal features of Transition Anatomy and their respective intervals.}
	\begin{tabularx}{\columnwidth}{l X}
		\toprule
		\textbf{Feature }& \textbf{Interval} \\
		\midrule
		\textit{PreDS} & $[L_\alpha(k_d),DS_\alpha(k_d)]$  \\
		\dhline
		\textit{DoubleSig} & $[S_\alpha(k_d),S_\phi(k_d)] \cap [S_\alpha(k_r),S_\phi(k_r)]$  \\
		\dhline
		\textit{PreStage} & $[L_\alpha(k_r),S_\alpha(k_r)]$  \\
		\dhline
		\textit{DepSigOnly} & $[S_\alpha(k_d),S_\phi(k_d)] \cap [T_\alpha,T_\omega] - DoubleSig$  \\
		\dhline
		\textit{Retire} & $[S_\phi(k_d),L_\omega(k_d)] \cap [T_\alpha,T_\omega]$  \\
		\dhline
		\textit{DSOverlap} & $[DS_\alpha(k_d),DS_\omega(k_d)] \cap [DS_\alpha(k_r),DS_\omega(k_r)]$  \\
		\dhline
		\textit{RemSigOnly} & $[S_\alpha(k_r),S_\phi(k_r)] \cap [T_\alpha,T_\omega] - DoubleSig$  \\
		\dhline
		\textit{DsPreRem} & $[DS_\omega(k_d),L_\omega(k_d)]$  \\
		\dhline
		\textit{RemPreDS} & $[L_\alpha(k_r),DS_\alpha(k_r)]$  \\
		\dhline
		\textit{TotalDuration} & $[T_\alpha,T_\omega]$  \\
		\raisebox{2pt}{$\llcorner$} for ZSK & $[L_\alpha(k_r),L_\omega(k_d)]$ \\
		\raisebox{2pt}{$\llcorner$} for KSK & $[min(L_\alpha(k_r), DS_\alpha(k_r)),max(L_\omega(k_d),DS_\omega(k_d)]$ \\
		\bottomrule
	\end{tabularx}
	\label{tab:transition-anatomy-intervals}
\end{table}

\noindent \autoref{tab:transition-anatomy-intervals} describes these features, using our life cycle notation of involved keys presented in \autoref{fig:temporal-model-dnskey}.

Based on our anatomy, we are able to describe  arbitrary transitions of $n$ departing keys to $m$ remaining keys, using $\binom{n+m}2$ pairwise $1:1$ transitions.
Our measurements of operational zones over 15 years indicate that this distinction is important.
For example, consider a zone of $n$ keys (with 
possibly different initial inception times) and $1$ to $n$ keys in use to sign data.  If that zone transitions to $m$ keys 
(where $1$ to $m$ are used to sign data), a number of unknowns arise: 
Which key(s) rolled over to which other keys?  Did all the departing keys roll over 
to all of the remaining keys?  If some keys persisted, are they (partially) replaced, as well?

Figure~\ref{fig:tot-card-chng} clearly
shows that while the majority of observed key transitions change by one key (gaining or losing), many transitions cause alterations of  $|m - n| > 1$. 
These fine granular observations of real deployments illustrate why additional expressiveness is needed, and why 
many of the previous discussions and characterizations of ``rollovers'' in the literature apply 
only to cases in which a single \emph{active} key is rolling over to another single active key.

We extend these discussions by observing that if more than one key is added or departed at the same time, these 
are multiple concurrent transitions at the same timestamp.  The intuition here is that no single departing key is measurably 
more pivotal than another.  Thus, we define each departing key as transitioning to any of the keys that remain.  This 
definition of key transitions allows us to measure operational behavior and answer questions such as: How many active keys are in use?
When are transitions aborted or rolled back? 
When are secure delegations (from {\tt DS} records) correct? 
Or, does a zone remain secure (see \autoref{sec:sec-analysis}) as transition is ongoing.
In addition, 
our transition model measures the relative ages of (the remaining to departing) keys:
\emph{newer}, \emph{older}, or the \emph{same} age.
These semantics, although unmentioned in the RFCs, are useful in some process models (\eg~\cite{wang2014emergency} discussed below).

Our proposed anatomy is a fine-grained description of the atomic timing components that are necessary and sufficient 
to fully characterize key transitions.
Whether key transitions are being performed manually, part of a process, fully automatized, or result from unsupervised ad hoc changes in a DNSSEC zone, 
the diversity of their activities is concisely represented by this anatomy.

\subsection{Transition Classes}\label{sec:desire-lines}

Our transition anatomy allows for the precise reconstruction and description of any pairwise key transition.
Special care needs to be taken when \emph{characterizing} transitions abstractly, as RFC~5011~\cite{rfc5011} or RFC~7583~\cite{rfc7583} do for example.
We discretize the value of each transition feature~(see \autoref{tab:transition-anatomy-intervals}) instead of using their absolute values from empirical measurements.
Different combinations of the resulting discretized feature set then represent transition classes.
Transition classes allow us to  compare and to assess whether prescribed security guarantees are preserved while keys are changing (\eg by adhering to specifications such as RFCs).

To classify a given transition, we first calculate the interval for each of its features using empirical measurements, \eg $PreDS = [1618016262, 1618048662]$.  
We then discretize the features (1) through (9) based on whether
their interval widths (see \autoref{tab:transition-anatomy-intervals}) are
$< 0$, $= 0$, $> 0$, \eg $PreDS > 0$.
For intervals defined as intersections of other intervals, \eg \textit{DoubleSig}, the respective interval widths are always non-negative, whereas other measures can assume negative values, \eg the width of \textit{PreDS} interval can be negative when a KSK is securely delegated \emph{before} being signed and included in the \texttt{DNSKEY} RRset ($L_\alpha(k_d) \ge DS_\alpha(k_d)$).
For cases in which an interval is undefined, \eg measurements of {\tt DS} records for ZSKs that have no {\tt DS} record to measure, we use the \textit{N/A} placeholder.
In \autoref{sec:classification}, we will see that N/A cases do not have an impact on transition classes.

Discretization facilitates an empirically simple comparison of completely independent key transitions.
For example, if two keys in a transition are both observed to be signing data at the same time, their observed \emph{DoubleSig} interval width would be a finite value.  This would then be
discretized as $> 0$.
Two other keys, in another transition (possibly in another zone) would likely have different interval widths, but would be assigned to the same discrete classification value, and would thus enable comparisons between these transitions.
As a result, every pairwise key transition can be represented as an ordered set of discretized features.

\section{Security Analysis of Key Transitions}
\label{sec:sec-analysis}

\begin{figure}
	\centering
		\tiny
	\includegraphics{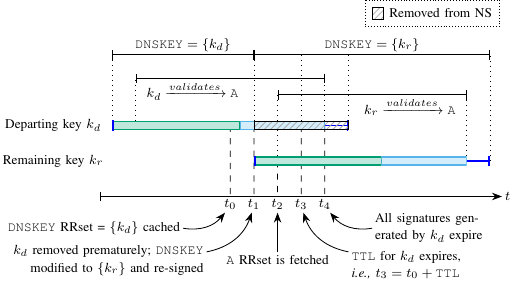}
	\caption{An example of how modifying and re-signing of an RRset (\texttt{DNSKEY} here) can cause failed validation in caching resolvers, while a valid signature over the RRset still exists.}
	\label{fig:failed-cache}
\end{figure}

In this section, we discuss the temporal constraints that must be satisfied during key transitions to avoid zone data to be invalidated as insecure or bogus~(see \autoref{sec:bg}).
Recall that a single record is considered valid if all RRsets that are necessary for the validation of that record are also valid at the time of verification.
To maintain this temporal requirement, each zone operator must ensure that at any point in time \one~secure delegations from the parent zone cover at least one KSK that signs the \texttt{DNSKEY} RRset, and \two all records are signed by at least one valid ZSK that is included in the \texttt{DNSKEY} RRset.
Breaking the first condition causes zone data to be invalidated as insecure, and ignoring the second leads to a bogus state.
This temporal requirement must also be maintained during key transitions, just as \texttt{DNSKEY}, \texttt{DS} RRsets, and \texttt{RRSIG}s are undergoing changes.
The main challenge, however, is not only to maintain a secure state on authoritative nameservers but to ensure that \one RPs can base validations on RRsets that are cached somewhere in the network and still carry valid signatures, and \two changes at primary authoritative nameservers are not instantaneously observed by relying parties (RPs)~\cite{RFC6781}.
Given a secure data state and a well-functioning RP, we consider a transition as secure if its modifications to keys, signatures, and secure delegations prevent a bogus or insecure state during and after the transition.
This definition also accounts for adversaries that can replay valid RRsets that might already be removed from the authoritative nameservers and purged from caches.
Our following security analysis considers the relation of a key to its generated signatures, and its relation to other keys and delegation signers.

\subsection{Temporal Relationship of a Key and its Signatures}
Based on our temporal model of DNSSEC keys~(see~\autoref{fig:temporal-model-dnskey}), the following condition applies for a key and its generated signatures: $L_\alpha \leq S_\alpha \leq S_\omega \leq L_\omega$.
This condition states that no signature validity period should precede or exceed the total (non-zero) lifetime of the key that generated it.
By extension, \texttt{DNSKEY} sets that have a valid signature may not be prematurely removed from authoritative nameservers without the risk of introducing inconsistencies through caching or replay attacks.
The following example illustrates such a situation: an RP caches the \texttt{DNSKEY} RRset of a zone and its \texttt{RRSIG}(s) at $t_0$.
Before the signature over the key set is expired, the authoritative nameserver replaces the ZSK at $t_1$, uses the KSK to generate a new signature over the \texttt{DSNKEY} RRset, and finally re-signs all other RRsets in the zone with the new ZSK.
At $t_2$, before either the signature or the \texttt{TTL} of the old \texttt{DNSKEY} expires, the recursive resolver fetches, for example, \texttt{A} records and their signatures generated with the new ZSK from the authoritative nameserver.
At this point, the resolver considers the cached key set as valid (\texttt{TTL} spans at $t_3$) yet fails to validate the retrieved RRset and concludes a bogus state (see~\autoref{fig:failed-cache}).
Even if the \texttt{TTL} was expired on all caches and the old \texttt{DNSKEY} RRset was purged at $t_2$, an adversary could still poison RP caches with the old key set and cause a failed validation.
This clearly illustrates that \texttt{TTL}s do not affect the foundational security analyses.

The scenario of failed validation because of bad timing becomes more severe in case of \texttt{DS} records.
When a newly signed \texttt{DS} record is fetched from an authoritative server but the RP uses a departing, cached \texttt{DNSKEY} for validation, a complete child zone becomes insecure.
This case, however, is more complicated because the only requirement to maintain the chain of trust is to have at least one active KSK securely delegated even if multiple KSKs are present and are actively signing the \texttt{DNSKEY} RRset.

\subsection{Temporal Relationship among Keys and Delegation Signers}
The second temporal constraint states that involved keys in a transition must have overlapping lifetimes ($TotalDuration > 0$) as changes on primary authoritative nameservers are not instantaneously observable by RPs~\cite{RFC6781}.
Changes to delegation signers must also account for additional delays 
because
as the 
operator of a 
zone can only request a modification to \texttt{DS} records to a parent zone, with no control over the timing when exactly changes are applied by that parent zone.

\begin{figure}%
	\tiny
	\centering
	\includegraphics{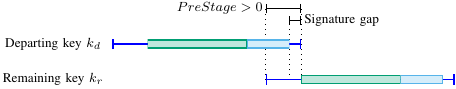}
	\caption{An example of improperly timed ZSK transition with pre-staged remaining key but belated activation causing bogus validation due to a signature gap.}
	\label{fig:bad-zsk-prestage}
\end{figure}

To avoid going bogus, a zone operator must guarantee the continuity of signatures over zone data.
As DNSSEC allows for multiple signatures over the same RRset at the same time, it suffices that for each RRsets in a zone to have at least one signature that is valid at all times.
Transitions that involve multiple keys must make sure that for any RRset signed by a departing key, a remaining key exists which signs that RRset no later than its signatures expire.
The most basic solution for this is to activate keys, \ie create \texttt{RRSIG}s, as soon as keys are introduced in the zone ($PreStage = 0$).
This way, any active departing key can stop generating new signatures (runout, \ie $Retire > 0$) and be removed after its generated signatures are all expired ($L_\omega \ge S_\omega$).
This approach, however, expands the zone size as multiple valid signatures are present simultaneously during the transition.
To address this, keys can be introduced to a zone and start signing later ($PreStage > 0$) to reduce the time window of overlapping signature.
Here, the downside is the more challenging timing in making sure that there is no signature gap for any available RRset.
\autoref{fig:bad-zsk-prestage} depicts an example where the remaining key is pre-staged but activated too late, thus, leaving a gap during which zone data cannot be validated. 

With KSKs, additional care must be given to the timing of \texttt{DS} records in order to maintain the chain of trust during the transition and avoid going insecure.
In details, this depends on how the KSK is being transitioned: if $PreStage > 0$, the new \texttt{DS} record can be added to the parent zone \emph{after} the key is included ($RemPreDs > 0$) but no later than expiration date of previous DS, while the old delegation can be removed before the departing key expires ($DSPreRem > 0$), yet no later than its runout period ends.
This approach can be used to minimize the interaction between the zone operator and its parent zone by combining the request to remove the old delegation and adding the new one ($DSOverlap = 0$).
If $PreStage = 0$, any combination of $RemPreDS$ and $DSPreRem$ that does not cause a gap between delegation signers ($DS_\alpha(k_r) \leq DS_\omega(k_d)$) can be applied.
This implies that when all signatures of the departing key expire, one remaining active key must still be securely delegated.

\section{Monitoring System and Data Corpus}\label{sec:corpus}

In this section, we introduce the DNSSEC measurements
taken from our monitoring system~\cite{secspider-web} that we used to evaluate key transitions in the wild.
We give an overview of our monitoring system, describe its operational aspects, and discuss the characteristics of the resulting data corpus, which covers the first 15~years of the global DNSSEC rollout.

\subsection{Monitoring System}

Our monitoring system collects DNSSEC records ({\tt DNSKEY}s, {\tt DS}es, {\tt RRSIG}s) alongside other types of RRsets, network PMTU measurements, name server versions, and many other relevant measurements by polling \emph{all} of \emph{every} zone name servers (as specified by both the zone and predecessor {\tt NS} records) from distributed vantage points across the Internet~\cite{secspider-acsac10, secspider-web}.
This comprehensive polling is a critical feature for observing key transitions with complex process models such as those specified in RFC 8901~\cite{rfc8901}.

In order to capture the holistic status of the global DNSSEC deployment, we broadly define zones as being \emph{DNSSEC-enabled} if they have deployed one or more {\tt DNSKEY} records.
The set of DNSSEC zones in this corpus was learned from proactive crawling of online sources, {\tt NSEC}-walking~\cite{kolkman2009dnssec, secspider-acsac10}, user-submissions, selected top-lists (Alexa Top One Million~\cite{alexa}, the Majestic Million~\cite{majestic}, and the Cisco Umbrella Popularity~\cite{umbrella}), and other techniques.
The development of the number of monitored zones throughout the years is depicted in \autoref{fig:zone-cdf}, and the detailed numbers are presented in Appendix~\ref{app:zone-count}.

\begin{figure}
	\includegraphics[width=\columnwidth,keepaspectratio]{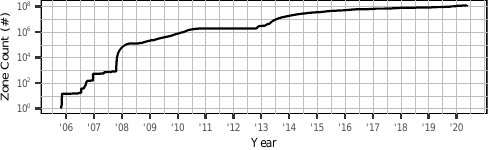}
	\caption{Total number of monitored DNSSEC-enabled zones.}
	\label{fig:zone-cdf}
\end{figure}

As our automatic discovery methods might catch non-production zones, \ie secure zones which are deployed for testing purposes, we apply a conservative heuristic to keep only production zones. First, all zones under \texttt{arpa} TLD are marked as production, then any zone that points to an active web or mail server is added to an include-list~\cite{Osterweil:2008:QOS:1452520.1452548}.
The remaining zones
are considered as non-production.
This way, we keep only zones for which we have positive indications of actually being in production.
It should be noted that since there is no systematic mechanism to discover all zones in the global DNSSEC, measurements can be subject to sample bias; nonetheless, we argue that our measurements of over 9.5 million DNSSEC zones at the time of this writing represent a relevant set on which to measure general behaviors of key transitions in the wild.

Our monitoring system polls DNSSEC zones concurrently in four ways:

\begin{enumerate}
	\item The monitoring system repeatedly queries all zones from each of its remote polling locations.  
	These full measurements poll hundreds of zones concurrently, and 
	when finished, the system starts again.
	As the number of DNSSEC zones has increased, the full-corpus polling period has grown from taking days to taking weeks,
	making the periodicity of polling a varying interval.
	\item
    In parallel,
	it polls 
    the DNS Root zone and all Top-Level Domains (TLDs) once per day, which is half the TTL period of those zones.
	\item
	In addition, it polls popular DNSSEC-enabled domains from our top-list collection every two days.
	\item
	Finally,
	the  interactive website of the monitoring system allows users to poll any zone on-demand. 
\end{enumerate}

Noteworthy, though, is that zones that delegate to others are implicitly queried and measured multiple times beyond individual schedule.
For example, the {\tt .com} zone is not only monitored once but its keys and their \emph{usages} are also 
re-observed for every delegated domain below, in order to assess {\tt NS} and {\tt DS} records.
For example, the root zone is measured over 1,300 times every day, once for every Top-Level Domain polled.
This greatly increases the frequency of observation, albeit in an aperiodic way.

Our monitoring system has occasionally undergone migration to new hosting infrastructures, database backups, and other operational maintenance events during the 15 years, leaving irregular gaps in our data corpus.
We control for gaps in our methodology, for details see Section~\ref{sec:methodology}.

\subsection{Data Corpus}

This work makes use of data from October 1,~2005 to August 31,~2020.
The resulting data corpus encompasses over 30.8~billion DNSSEC measurements from 9,535,615~DNSSEC-enabled zones with 35,882,395 distinct {\tt DNSKEY}s. 
We observed 58,193,197 points in time when keys were either added or removed from zones. Of those changes, 17,965,575 key transitions %
were detected (see \autoref{sec:key-trans-anat}) and analyzed, which we will discuss in Section~\ref{sec:meas}.

Our data corpus spans several events that are notable for the global deployment of DNSSEC.
First,
multiple Top-Level Domains (TLDs) such as  {\tt .com}, {\tt .edu}, and 145 country-code TLDs~(ccTLDs) deployed DNSSEC for the first time.
Second, the announcement of a crucial large-scale DNS cache poisoning attack vector called the Kaminsky Attack~\cite{kaminsky2008black,cert2008vulnerability}, 
whose remediation was publicized to be the deployment of DNSSEC.
Third, the DNS Root zone enabled and rolled out DNSSEC for the first time using a Deliberately
Unvalidatable Root Zone~(DURZ) key.  Fourth, in 2010, the DNS Root zone performed its first ever Key Signing Key (KSK) transition, from the DURZ key to the 2010 KSK.  
Fifth, over 1,200 DNSSEC-enabled new generic Top-Level Domains were added since 2013.
Finally, the Root KSK was transitioned for the second time ever in 2018, after being started and paused during 2017.
\autoref{fig:timeline} depicts a number of these notable events and incidents 
and by overlaying the measurement periods of related work
and this study, shows this work has a uniquely complete longitudinal dataset to draw conclusions from.

\section{Methodology}\label{sec:methodology}
The initial step in our methodology is to reconstruct the lifetime of \texttt{DNSKEY}s according to our discrete measurements and in accordance with our proposed temporal model (see \autoref{sec:key-trans-anat}).
This is a deceptively challenging step because when keys are provisioned into zones there are no semantics to express (or for zone administrators to even know) life cycle information.
In general, key lifetime management and  changes such as re-signings or deployment of new keys may occur at varying times between polling cycles.
This will lead to 
gaps in time between three events: \one when these changes happen on the authoritative name servers, \two  when we poll the zones, and \three when we observe them in use. This
may lead to
changes that a measurement system completely misses regardless of the polling frequency.
A simple example is the replacement of a \texttt{DNSKEY} multiple times between two polls.

We now give a brief overview of our methodology and then present the key building blocks in more detail.

\begin{figure}
	\centering
	\includegraphics{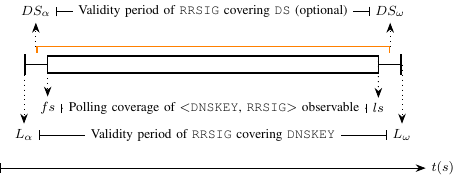}
	\caption{Visualization of an observable for a given $<$\texttt{DNSKEY},\texttt{RRSIG}$>$ tuple.}
	\label{fig:dnskey-ds-observable}
\end{figure}

\subsection{Overview}

Our first step in reconstructing the complete and continuous lifetime of keys from our discrete measurements is to infer a so-called \textit{observable} for any unique $<$\texttt{DNSKEY}, \texttt{RRSIG}$>$ tuple and any respective secure delegation period (if applicable) while preserving points of time when this specific observable was seen during our measurements.
If the lifetime of a key is extended by re-signing (\ie generating a new \texttt{RRSIG} over the same \texttt{DNSKEY}), our methodology creates a new observable~tuple.
In this sense, an observable can be considered as a piece of evidence based on a given \texttt{RRSIG} denoting that a key has been valid from the inception to expiration dates of that \texttt{RRSIG}.
\autoref{fig:dnskey-ds-observable} depicts \emph{one} observable for key $k_x$. The lifetime of this single observable duration $[L_\alpha, L_\omega]$ is calculated from the signature covering $k_x$; 
its delegation period $[DS_\alpha, DS_\omega]$ is inferred from the signature in the parent zone covering \texttt{DS} record(s).  $fs$ and $ls$ denote the times when this single observable was first and last seen during the measurements.
Note that $fs$ is recorded once per observable, and never changed again, while $ls$ is updated when the same $<$\texttt{DNSKEY}, \texttt{RRSIG}$>$ tuple is encountered again in a subsequent measurement.

Individual observables expand the information about a key lifetime from a discrete snapshot to a continuous timeline within the validity period of its covering \texttt{RRSIG}.
Such extrapolation of observables, however, might still leave some \textit{gaps} in our continuous lifetime model.
Furthermore, other life cycle information, such as key usage (see \autoref{fig:temporal-model-dnskey}), cannot be inferred from \texttt{RRSIG}s over keys but must be measured from signatures \emph{generated} by those keys when they are in use.
To address this we introduce a novel three-step methodology that we call \emph{Bridging, Busting,} and \emph{Binding}:

\begin{description}[\IEEEsetlabelwidth{Bridging}]
	\item[\textbf{Bridging}] Extend observables by filling in measurement gaps with place-holder observables, which we call bridging ``ghosts''.
  \item[\textbf{Busting}] Use collected evidence, \eg \texttt{RRSIG}s over non-\texttt{DNSKEY} records, to remove (or ``bust'') incorrect ghosts.
	\item[\textbf{Binding}] For any given \texttt{DNSKEY} combine remaining contiguous observables into a continuous holistic life cycle model.
\end{description}

This process extends sets of observables into full key life cycles and builds a basis to calculate related statistics such as signing frequency, measure management errors, and compute other aggregate behaviors.
It also accounts for gaps in our data corpus.

\begin{figure}
	\begin{subfigure}{\columnwidth}
		\centering
		\includegraphics{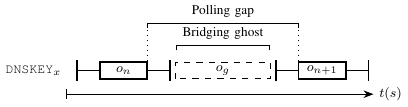}
		\caption{Bridging between observables of the same \texttt{DNSKEY}}
		\label{fig:gap-singlekey}
	\end{subfigure}
	\begin{subfigure}{\columnwidth}
		\centering
		\includegraphics{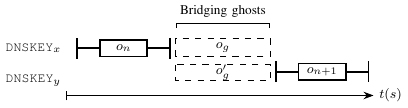}
		\caption{Bridging between observables of different \texttt{DNSKEY}}
		\label{fig:gap-multikey}
	\end{subfigure}
	\caption{Filling in observation gaps with bridging ghosts.}
\end{figure}

\subsection{Bridging}
Bridging begins by time-ordering observables for each zone.
If the maximum time of a single observable (\ie $max(L_\omega,ls$)) is less than the minimum time of the next observable (\ie $min(L_\alpha,fs$),
then a \textit{ghost record} is inserted, which exactly covers the missing time.
Note that a ghost does not necessarily fill the whole gap between two polling cycles, but only the time for which no temporal information can be inferred (see \autoref{fig:gap-singlekey}).
If the ghost observable covers a gap between the observables of the same \texttt{DNSKEY} (but different \texttt{RRSIG}s), then the ghost observable proposes that the key existed continuously during the gap,
as shown in Figure~\ref{fig:gap-singlekey}.
If, on the other hand, the gap exists between observables of two different \texttt{DNSKEY}s, then the process cannot know precisely when the older \texttt{DNSKEY$_x$} stopped being present and the newer \texttt{DNSKEY$_y$} was deployed, if or when they may have overlapped, or if there was a period of no key(s) being present.
Therefore, at this stage, a bridge is \emph{temporarily} used, whereby a trailing ghost for \texttt{DNSKEY$_x$} is inserted until both the next observable and a leading (overlapping) ghost for \texttt{DNSKEY$_y$} are inserted, starting just after the previous observable, as seen in Figure~\ref{fig:gap-multikey}.
Ghosts represent optimistic assumptions about consistency between observations, but in the next phase we \emph{bust} ghosts if additional evidence proves they are incorrect or need to be adjusted.

\subsection{Busting}
Ghost observations model place-holders of inferred data that may have existed between the data we observed.
For each zone, after the initial optimistic Bridging phase, our process begins to examine keys in relation to each other and incorporating additional evidence to
detect if a ghost assumption can be refuted (and thereby \emph{busted}).
For this, additional measured data (such as {\tt NS}, {\tt SOA}, {\tt A}, and associated {\tt RRSIG}s) allow the process to determine if, or how, a ghost should be busted.
For example, if a ghost bridges a key, but another key was seen during that time, we can determine the ghost-key was not present for the period when the other key was seen.  
The ghost is, then,   busted by truncating it to the time interval(s) that are not covered by the other key.
Alternately,
if ghosts for two keys overlap between a transition (see Figure~\ref{fig:gap-multikey}), information about which of them was able to verify signatures over other measured data
is used to determine when one key was removed and the other was present, and truncate ghost overlaps accordingly.
Overall, ghosts may be truncated if a zone's data
was observed and the key was absent, and affirmed if a ghost's key was seen to be signing other records.
Ghosts which may have been causes by measurement outages are removed altogether.
Gaps caused by outages are distinguished through their relatively long duration.
For our purposes, we take the yearly statistics of our measurement system crawl times, \ie a complete round of polling all zones, to bust gaps caused by outages: for any given year all gaps that are larger than the mean crawl time plus four standard deviations are considered as an \emph{outage gap}, are busted and removed.
While the removal of long-ghosts
could result in missed key transitions and key management behaviors, not removing them could alternately 
enshrine inaccurate assumptions.

\begin{figure}
	\centering
	\includegraphics{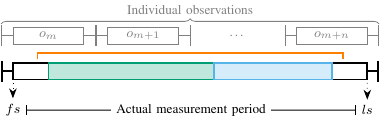}
	\caption{Observations are bound together to find minimum and maximum timestamps for inception, expiration, first seen, and last seen.  Measuring bounds also seek  evidence of when keys were used to generate signatures (green region) and how long their signatures were valid for (blue region).}
	\label{fig:bound}
\end{figure}

\subsection{Binding}
For every key, all the contiguous observations (including unbusted ghosts)
are used to \emph{Bind} observables into one single continuous key, as seen in Figure~\ref{fig:bound}.  
Bound keys describe the
life cycle of a DNS key in terms of our temporal model (\autoref{fig:temporal-model-dnskey}) with two additional features.
The first timestamp that the key was measured ($fs$) and the last time a key was observed ($ls$).
Here, all the observed {\tt RRSIG} records that could be verified by each key, regardless of which DNS record type they covered, 
are used to 
quantify whether keys were in active use.  The inception timestamp of each {\tt RRSIG} is used as evidence to indicate when a key pair was signing data ($[S\alpha,S\phi]$). The expiration of
that signature specifies the run-out, or the duration that a key's data was verifiable while no new signatures were detected ($[S\phi,S\omega]$).

\section{Classifying Keys and Transitions}
\label{sec:classification}

Using our model of continuous life cycles for {\tt DNSKEY}s,
we move on to classification of \texttt{DNSKEY} life cycle states and key rollovers and more complex key transition types.

\subsection{Key Life Cycle Classification}
\begin{figure}
	\begin{center}
		\includegraphics{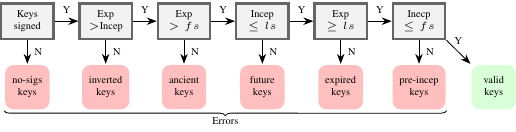}
		\caption{Classification of key life cycle management states. Each state results from an observable property (rectangle) that is either fulfilled (Y) or not fulfilled~(N). Properties include expiration (Exp) and inception (Incep) dates, as well as first ($fs$) and last seen ($ls$) of operational usage.}
		\label{fig:taxonomy}
	\end{center}
\end{figure}

We present our novel classification scheme of measurable errors in key life cycle management in Figure~\ref{fig:taxonomy}.
Based on this model we observe previously undetected errors in live deployments, which we explain in more detail in \autoref{sec:meas}.

This classification scheme defines six types of key life cycle management errors depending on when a key is introduced in the zone and how inception and expiration dates of the respective \texttt{RRSIG} are defined.
For example, if the expiration predates the inception date we classify the key as \emph{inverted}, or if key was observed before its actual inception date, we label it as a \emph{future} key.
Avoiding these errors is a necessary (yet insufficient) precondition for \emph{valid} cryptographic protections.

\renewcommand{\arraystretch}{1.2}
\begin{table*}[t]
	\footnotesize
	\caption{
		Mapping of key transition processes specified in DNSSEC~RFCs and other literature to Transition Anatomies. 
		In each row, gray cells show the mandatory features that must be fulfilled to map a transition to a predefined class, whereas the other cells describe soft constraints, which only cause warnings if not followed exactly. Cells showing ``--'' indicate wildcards.}
	\label{tab:feature-vector-w-emerg}
	\setlength\tabcolsep{3.1pt}
	\begin{tabular}{l l c c c c c c c c c c}
		\toprule
		& & \multicolumn{10}{c}{\textbf{Features of the Transition Anatomy}} \\
		\cmidrule{3-12}
		& \textbf{Transition Class} & \emph{PreDS} & \emph{ DoubleSig} & \emph{ PreStage} & \emph{ DepSigOnly} & \emph{ Retire} & \emph{ DSOverlap} & \emph{ RemSigOnly} & \emph{ DSPreRem} & \emph{ RemPreDS} & \emph{TotalDuration} \\
		\midrule
		& \emph{ZSK Pre-Pub} & -- & $= 0$ & \cellcolor{lightgray} $>0$ & $> 0$ & $> 0$ & -- & $> 0$ & -- & -- & --\\
		\dcline{2-12}
		& \emph{ZSK Double-Sig} & -- & \cellcolor{lightgray} $> 0$ & \cellcolor{lightgray} $= 0$ & $= 0$ & $= 0$ & -- & $= 0$ & -- & -- & --\\
		\dcline{2-12}
		& \emph{KSK Double-DS} & $< 0$ & $= 0$ & $= 0$ & $= 0$ & $= 0$ & \cellcolor{lightgray} $> 0$ & $> 0$ & \cellcolor{lightgray} $< 0$ & \cellcolor{lightgray} $< 0$ & --\\
		\dcline{2-12}
		& \emph{KSK Double-KSK} & $> 0$ & $> 0$ & $= 0$ & $= 0$ & $> 0$ & $= 0$ & $> 0$ & \cellcolor{lightgray} $> 0$ & \cellcolor{lightgray} $> 0$ & --\\
		\dcline{2-12}
		\rotatebox{90}{\rlap{Based on RFCs}}
		&
		\emph{KSK Double-RRset} & $> 0$ & $> 0$ & $= 0$ & $= 0$ & $> 0$ & $= 0$ & $> 0$ & \cellcolor{lightgray} $\ne 0$ & -- & --\\
		\dcline{2-12}
		&
		\emph{KSK Update-TA} & -- & -- & -- & -- & -- & -- & -- & -- & -- & \cellcolor{lightgray} $\ge 30d + AvgSig$\\
		\midrule
		& \emph{KSK Emergency 2} & -- & \cellcolor{lightgray} $= 0$ & \cellcolor{lightgray} $= 0$ & $= 0$ & \cellcolor{lightgray} $= 0$ & \cellcolor{lightgray} $= 0$ & $> 0$ & -- & \cellcolor{lightgray} $\ge 0$ & --\\
		\dcline{2-12}
		& \emph{KSK Emergency 3} & -- & $= 0$ & \cellcolor{lightgray} $> 0$ & -- & $> 0$ & -- & -- & $> 0$ & \cellcolor{lightgray} $> 0$ & extended \\
		\dcline{2-12}
		& \emph{ZSK Emergency 2} & -- & \cellcolor{lightgray} $= 0$ & \cellcolor{lightgray} $= 0$ & \cellcolor{lightgray} $= 0$ & \cellcolor{lightgray} $= 0$ & -- & $> 0$ & -- & -- & -- \\
		\dcline{2-12}
		\rotatebox{90}{\rlap{Based on \cite{wang2014emergency}}}
		& \emph{ZSK Emergency 3} & -- & -- & \cellcolor{lightgray} $> 0$ & -- & -- & -- & -- & -- & -- & extended\\
		\bottomrule
	\end{tabular}
\end{table*}
\renewcommand{\arraystretch}{1.0}

\subsection{RFC-Based Classification}
The IETF has specified many aspects of how DNSSEC zones and data should be configured and maintained across
numerous RFCs~\cite{rfc4641,RFC6781,rfc5011,rfc7583}.  Among those are several processes that model ways in which {\tt DNSKEY}s should be
rolled over.  In RFC~5011~\cite{rfc5011}, Trust Anchor (TA) rollover is specified for zones
whose predecessor zones do not securely delegate {\tt DS} records.  Additionally,
in RFC~7583~\cite{rfc7583}, processes are described for how zone administrators should
transition their ZSKs and KSKs.

First we characterize RFC guidelines in terms of our transition model and then investigate how transitions in the wild conform to these guidelines (in \autoref{sec:meas}).
We classify deviations as either warnings, \ie the behavior does not strictly follow the RFC guidelines, yet, does not disturb validation; or as errors, which render the validation as bogus or insecure.
It should be noted that while conventional IETF parlance (\eg \texttt{MUST} vs. \texttt{SHOULD} in RFC~8174~\cite{RFC8174}) often makes this distinction explicit, the measurable quantities in RFC-7583~\cite{rfc7583} do not use \texttt{MUST, MAY, SHOULD}, etc.  We, therefore, semantically assign these values based on the overall processes.
For example, if a transition was specified as needing to have \emph{Retire}~$>0$,
but it was 0, this would only be a warning because the key transition would still allow keys to verify data for a zone.
In contrast to this, if a KSK must be present \emph{before} a {\tt DS} record in order to let resolvers securely verify the KSK during 
a \emph{Double-DS} transition, then \emph{PreDS} $\le 0$ is a critical~error.

\paragraph{RFC~5011 (Update of Trust Anchors)}
This transition is specified by a 
Finite State Machine with timers and a rigorous process model~\cite{rfc5011}.  
The process-model, however, is specified from a resolver's perspective (\ie
timers that a resolver should set internally), which do not always directly correlate to 
observable timing of {\tt DNSKEY}s in authoritative zones.  Additional guidance~\cite{draft-ietf-dnsop-rfc5011-security-considerations} was
written for authoritative zone administrators,
which lends itself more directly to being measured.
These two publications~\cite{rfc5011, draft-ietf-dnsop-rfc5011-security-considerations} define the specification for RFC~5011 transitions 
as those whose \emph{TotalDuration} $\ge 30 days + min(15d,\frac{1}{2}\times TTL)$,
where
30~days are specified in~\cite{draft-ietf-dnsop-rfc5011-security-considerations} and 
the key's Time-To-Live (TTL) is specified as an additional component of the period.  
Additionally, 
those keys
that are being removed must be revoked \emph{and} be used to sign their own revoked {\tt DNSKEY} set.
In order to be conservative and permissive, we model the upper-bound (\ie $max$ instead of $min$) from 
each key's own average signature period.
We conservatively modeled those zones 
that did adhere to RFC~5011 timing, but \emph{did not}
revoke as still being RFC~5011 compliant, but flagged them with a warning.

\paragraph{RFC~7583 (DNSSEC Key Rollover Timings)}
RFC~7583 specifies the process models for both ZSKs and KSKs, and they are more stringent.
For ZSK, these
transitions are defined as either being ``Pre-Publication'' or ``Double-Signature''; and for KSK as one of ``Double-DS,'' ``Double-KSK,'' ``Double-RRset,''
or ``Double-RRSIG''.
Note that the final type is identified as unrealistic in the RFC, and not
even fully described there; therefore, we also omit it here.

\begin{figure}
	\includegraphics{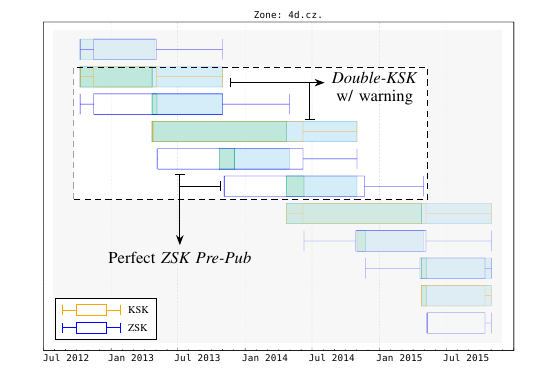}
	\caption{An example from our measurements showing how \texttt{4d.cz.} zone performs a perfect \emph{ZSK Pre-Pub} transition and a \emph{Double-KSK} transition with warnings (\emph{Double-Sig} $\ngtr 0$ and \emph{Retire} $\ne 0$).}
	\label{fig:transition-example}
\end{figure}

\autoref{tab:feature-vector-w-emerg} summarizes how RFC-based classifications are translated into our transition anatomy from \autoref{sec:key-trans}.
An illustrative example is provided in \autoref{fig:transition-example}, depicting an excerpt of transitions for a representative zone, \texttt{4d.cz.}. As is shown, the zone performs a correct ZSK pre-publication transition while (at other times) performing double-KSK transitions \emph{with warnings} (see \autoref{tab:feature-vector-w-emerg}).
Other examples of key transitions are visualized in Appendix~\ref{app:wild}.

\subsection{Non-IETF Prescriptions: Emergency Key Transitions}
In addition to the above guidance from RFCs, other prior work by Wang~\etal~\cite{wang2014emergency} proposed an approach 
for conducting emergency key transitions.  There, the authors 
present 10~candidate process prescriptions for emergency key transitions, which are
distinct from conventional RFC guidance.
It is noteworthy that 
this prior work considers keys transitions to stand-by keys which are envisioned to be perpetually present before the emergencies.
This, therefore, does not precisely specify the \emph{TotalDuration} threshold, but suffices to make the ``extended'' duration feature a necessary detectable discriminator in these emergency
transitions classification.
Table~\ref{tab:feature-vector-w-emerg}
shows how
the timing constraints detailed in that work can be used to classify and detect these events specified by our anatomy and transition methodology, just as with RFC
guidance.
Due to space limitations, we only include Emergency~Transitions~2 and 3.

Additionally, aspects of that work use features of our anatomy and our model of key life cycles 
that the RFC processes did not
(specifically, the TotalDuration and relative ages).
For example, the specification requires remaining keys to be \emph{newer} (see \textit{relative ages} in~\autoref{sec:key-trans-anat}), \emph{ZSK~Emergency~2} mandates the feature-set of the behavior-based classification \emph{Multi-Signatures} (described below, in Section~\ref{sec:behavior-class}), and \emph{ZSK~Emergency~4} requires \emph{Cutover}s.

In summary,
using the prescriptions of prior work, we detected 49,894 emergency ZSK transitions (20,919 that 
were transitioning \emph{back} or \emph{aborting} transitions to older or same age keys), 
and 1,780,984 emergency KSK transitions (149,406 transitioning \emph{back}  or \emph{aborting} to older or same age keys).

\subsection{Behavior-Based Classification}\label{sec:behavior-class}
In 
our behavior-based classification,
we classify key transitions as being 
``Multi-Signatures,''
``Co-Present,''
``Cutovers'' (with degrees of certainty), or
``Unknown''.
This classification approach uses more
holistic considerations of all keys in a transition (not just pair-wise) while ignoring the relationships to {\tt DS} records.
Here, key transitions were classified based on \one the
type and the total count of overlaps they had with \emph{all} other \emph{active} keys, and \two whether they were used to
verify zone data during their transitions. 

\setcounter{figure}{13}
\begin{figure*}[b!]
	\centering
	\begin{subfigure}{\columnwidth}
		\centering
		\includegraphics{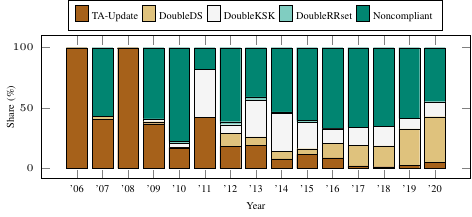}
		\caption{Key Signing Keys (KSK)}
		\label{fig:used-ksk-rfcclass}
	\end{subfigure}
	\begin{subfigure}{\columnwidth}
		\centering
		\includegraphics{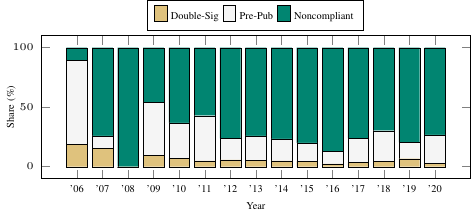}
		\caption{Zone Signing Keys (ZSK)}
		\label{fig:used-zsk-rfcclass}
	\end{subfigure}
	\caption{RFC-based classification of key transitions for in-use KSK and ZSK.}
	\label{fig:trans-rfcclass-per-year}
\end{figure*}

\setcounter{figure}{12}
\begin{figure}
	\centering
	\includegraphics{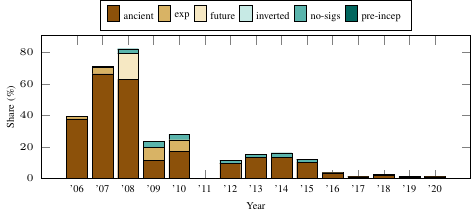}
	\caption{Classified key management errors.}
	\label{fig:yearly-err-perc}
\end{figure}
\setcounter{figure}{12}

When more than one key was seen to be actively in-use (verifying signatures) at any given time, we classified the transition as
a ``\emph{Multi-Signature}'' transition.  This indicates that redundant verification existed for data in a zone during
the transition.
If, on the other hand, keys were observed to have overlapped, but we did not observe \emph{any} of them in use, we classify
the transition keys as being ``\emph{Co-Present}.''  This classification does not represent evidence that keys went unused,
but indicates that our observations did not detect usage.  
Conversely, transitions are classified as ``Cutovers'' when a single key was observed to transition 
(or cut-over) to another single key, and they were seen to have been used to verify signatures.  
This type of transition depends heavily on measurement frequency (whereby the
longer the gaps in observations, the more information is estimated from ghost records).  Because of this, 
we further sub-classify Cutovers as ``Cutover,'' ``Likely-Cutover,'' or ``Candidate-Cutover.''  The differences
between these are based solely on how much usage (\ie active signing) was directly observed.  
If active signing (the inception of {\tt RRSIG}s) was observed for the departing key right up until the
remaining key began to be used to verify {\tt RRSIG}s (and not after), then we classify this as a ``\emph{Cutover}.''
If, on the other hand, we did not observe new signatures, but the signature run-out (see \autoref{fig:bound})
of the departing key overlapped with the signatures of the remaining key, we classify the transition as a ``\emph{Likely-Cutover}''.
Finally, 
if the departing key was used at any point, and later the remaining key began being used with no other signatures seen in the time-gap, 
we classify this as a ``\emph{Candidate-Cutover}.'' The goal of these distinctions is to make our sub-classification results useful, but to also preserve their transparency.

\section{Longitudinal Analysis of \\ 15 Years of DNSSEC Key Transitions}\label{sec:meas}

In this section, we present the results from answering the following three questions: 
\one
Do keys involved in transitions follow proper life cycle management policies? 
\two  How does the anatomy of key transitions in practice compare with RFC guidelines specifying rollovers?
\three 
Which characteristics are popular in key transitions beyond key rollovers in terms of our behavior-based classification?

\subsection{Key Life cycle Management}
The atomic management of the keys can end in errors, independent of transitions of other keys.
Using the key life cycle management classification described in Section~\ref{sec:methodology}, we broadly examine the rates at which individual {\tt DNSKEY}s follow proper key management life cycles. Our results are summarized in \autoref{fig:yearly-err-perc}.
Key life cycle management errors were clearly highest during the early years of the DNSSEC global rollout.
At that time, the tools that supported DNSSEC were in their nascent stages of maturity, which is very likely the core reason for larger rates of key life cycle management errors.
This figure also illustrates that the error rates for keys were highest in 2008, just as the discovery rate of new DNSSEC zones surged and more than doubled the size of the global deployment in just a few months.
This all correlated in time with publicity to remediate the newly announced Kaminsky vulnerability~\cite{cert2008vulnerability} by deploying DNSSEC, which could have also correlated with inexperienced operators making operational errors in a rush to secure their deployments.

\subsection{Conformance to RFC Guidelines}
\paragraph{KSK and ZSK Transitions in the Wild}
\autoref{fig:trans-rfcclass-per-year} summarizes RFC-based classifications of both measured KSKs and ZSKs per year. Any transition that could not be classified as RFC-compliant is marked as \emph{noncompliant}.
This figure
confirms a common expectation but also reveals new insights.
Unsurprisingly,  in the early DNSSEC deployment KSKs were transitioned according to RFC~5011 (denoted here as ``TA-Update'').
This was necessarily the case because at that time there were very few registrars and also few parent zones that were able to offer secure delegations. 
Most Top-Level Domains did \emph{not} deploy DNSSEC and were, therefore, unable to securely delegate. 

\setcounter{figure}{15}
\begin{figure*}[b]
	\centering
	\begin{subfigure}{\columnwidth}
		\centering
		\includegraphics{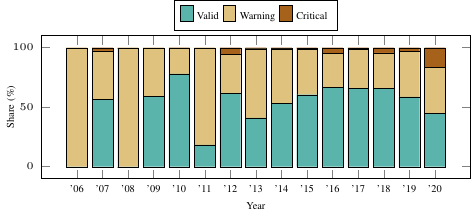}
		\caption{Key Signing Keys (KSK)}
		\label{fig:ksk-trans-err-per-year}
	\end{subfigure}
	\begin{subfigure}{\columnwidth}
		\centering
		\includegraphics{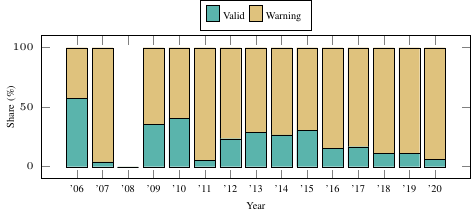}
		\caption{Zone Signing Keys (ZSK, excluding type noncompliant)}
		\label{fig:zsk-trans-err-per-year}
	\end{subfigure}
	\caption{Share of valid, warning, and error rates of in-use KSK and ZSK.}
	\label{fig:trans-err-per-year}
\end{figure*}

In 2008, operational advice was given to deploy DNSSEC to counter the Kaminsky vulnerability~\cite{cert2008vulnerability}.
Our data (see \autoref{tab:zone-count-year-source} and \ref{tab:raw-key-rfc-class}) indicate that the number of DNSSEC-enabled zones more than doubled during three months.
Based on our classification, it is clearly visible that subsequently more variety in key transition techniques appeared (see \autoref{fig:used-ksk-rfcclass}).
In 2009, the Double-DS transition technique was the most popular.
This technique, however, requires additional coordination between an authoritative zone and the operator of its parent zone.
Despite an increasing deployment over time, our results illustrate that the popularity of managing transitions based on  Double-DS and Double-KSK fluctuates.
The majority always conformed to either RFC~5011 or was ``noncompliant.'' 
This observation indicates that the RFC-specified key transition processes may not properly represent operational behavior.

Our results also exhibit an interesting discrepancy compared to related work.
In the longitudinal study of Chung~\etal~\cite{chung2017longitudinal}, no Double-DS transitions were reported.
However, under \texttt{.com} alone, we observed 17,126 unique zones performing Double-DS transitions during the time period of
their daily scans (from 2015-03-01 to 2016-12-31), and 7,256 transitions during the same period as their
hourly scans (from 2016-09-29 to 2016-12-31).
A concrete example of Double-DS transitions is \texttt{willemvanveldhuizen.com} (see \autoref{fig:willemvanveldhuizen.com-trans}).
Further understanding the discrepancies of these findings will be part of our future work.

\setcounter{figure}{14}
\begin{figure}%
	\centering
	\includegraphics{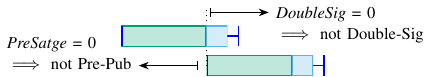}
	\caption{An example of RFC noncompliant transition performed among others by Root and \texttt{.com} zones}
	\label{fig:ex-noncomp-zsk}
\end{figure}
\setcounter{figure}{16}

\begin{figure*}%
	\centering
	\begin{subfigure}{\columnwidth}
		\centering
		\includegraphics{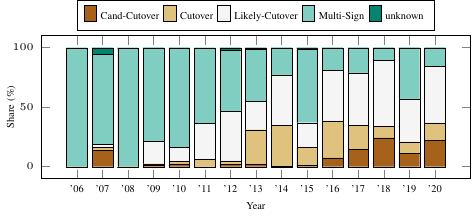}
		\caption{Key Signing Keys (KSK)}
		\label{fig:used-ksk-ssclass}
	\end{subfigure}
	\begin{subfigure}{\columnwidth}
		\centering
		\includegraphics{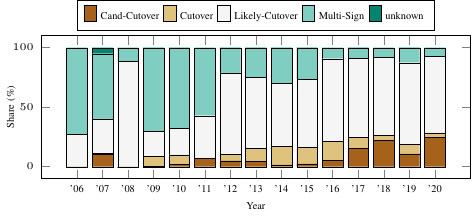}
		\caption{Zone Signing Keys (ZSK)}
		\label{fig:used-zsk-ssclass}
	\end{subfigure}
	\caption{Behavior-based classification of key transitions for in-use KSK and ZSK.}
	\label{fig:trans-ssclass-per-year}
\end{figure*}

In comparison to KSK transitions, an even larger
portion of ZSK transitions have constantly been nonconforming to RFC specifications (see \autoref{fig:used-zsk-rfcclass}).
While the ``Double-Signature'' alternative never accounted for more than 10\% of transitions, it is most noteworthy that between 2007 and 2020 (except for 2009) the majority of the observed ZSK transitions 
did \emph{not} conform to either prescribed mode of key transition.
Ignoring RFC~guidelines, however, does not necessarily mean that the 
zone 
verification would fail
during the transition.
\autoref{fig:ex-noncomp-zsk} depicts an example of such transition performed by prominent zones such as the Root and \texttt{.com} zones. 
This transition is neither Pre-Pub (\textit{PreStage} $\ngtr 0$), nor Double-Sig (DoubleSig $\ngtr 0$), yet at any time during the transition the RRs can be verified by either or both keys.
Here, we also see discrepancies with prior work~\cite{chung2017longitudinal}, which might be traced back to different approaches used in classifying transitions.

\paragraph{Implications on Robustness}
The warning and error rates for RFC-compliant transitions are notably different for KSKs and ZSKs, as depicted in \autoref{fig:trans-err-per-year}.
KSK transitions show more valid cases than ZSK transitions overall.
This corresponds also with  a higher share of transitions following RFC guidelines (see \autoref{fig:used-ksk-rfcclass}).
Later, between 2011 and 2020, critical error rates became even more prominent.
This could be the result of inherent management complexity in terms of timing and data for (longer) chain-of-trust-based transitions between authoritative zones ({\tt DNSKEY}s) and parent zones ({\tt DS} records).

For ZSKs, Figure~\ref{fig:zsk-trans-err-per-year} shows that the rate of warnings exceeded successes in every year, except in 2006,
when the DNSSEC deployment was just beginning, and zones were deploying DNSSEC for the first time.
In 2007, warning rates greatly exceeded success. 
In 2008, we observed only noncompliant transitions (see \autoref{tab:raw-key-rfc-class}), and from that year on the trend was increasing for success rates.
In 2011, 
an outage in our monitoring system resulted in fewer observed transitions, and the success rates of that smaller set may have been skewed.
success rates dropped precipitously, mainly due to 
From 2012, after a brief improvement, we also observe decreasing success rates.
This, however, again correlated with a large increase in the discovery rate of new DNSSEC-enabled zones.
During this time, the DNS Root was deploying DNSSEC and transitioning its KSK for the first time.
This
correlation between a large increase in newly deployed DNSSEC zones and increase in error rates suggest that these rates may also
be due to operators who were learning how to manage the security postures of DNSSEC in their DNS zones.

\subsection{Holistic Characteristics of Transitions}

Using the RFCs to classify key rollovers gives us a helpful start in understanding how to analyze key transitions.
However, to overcome the amount of noncompliant and high error rates, we use the alternate behavior-based classification scheme introduced in Section~\ref{sec:methodology}.

When applying our behavior-based classification to ZSKs that were seen to be in use, Figure~\ref{fig:used-zsk-ssclass} shows that 
almost every key transition observed for 15 years could be classified.
Only in 2007, 4.32\% transitions were classified as unknown; in all other years, $>$99\% of transitions were classified successfully.
The trend over the first 15 years was a shift from Multi-Signature, to cleanly cutting over from a departing key to a remaining key.

Figure~\ref{fig:used-ksk-ssclass} illustrates that all of these trends were similar in the KSK transitions.
A greater portion of the KSK transitions were Multi-Signature than in ZSKs, but the trend towards moving to Cutover (and the small incidence of unknown transitions) mimics the ZSK transitions.
This could be due to the same operational approach being used for both ZSKs and KSKs by zone operators, but the DNSSEC tools available have traditionally offered more comprehensive automation for ZSK transitions, since 
 they are managed within a single authoritative zone. By contrast, KSK transitions involve coordination with
the parent zone.

\section{Related Work}\label{sec:relwork}

The topic of DNSSEC key rollovers has recently appeared in the literature.  
In 2017, the DNS Root zone began publicity around its, then, intentions to transition its KSK for the second time (since DURZ, as discussed in Section~\ref{sec:corpus}).
At that time, van Rijswijk-Deij~\etal~\cite{10.1145/3123878.3131987} began the Root Canary project to track this specific transition.
Later,
M\"uller~\etal~\cite{muller2019rolling} created a related monitoring tool to
aid operators in successfully planning and conducting {\tt DNSKEY} rollovers.  More recently and maybe most closely related to our work, M\"uller~\etal~\cite{Muller:2019:RRR:3355369.3355570}
conducted an in-depth study of key life cycle management in the DNS Root zone.
Both of these results synergize with
the observations of our work: key life cycle management of DNSSEC zones is critical to their operational health.
Most importantly, and in contrast to prior work, we do not only consider key rollovers but provide a framework to analyze key transitions in general.

As DNSSEC deployment has grown, numerous studies have incrementally tracked its deployment.  Among the first measurement studies, Osterweil~\etal~\cite{Osterweil:2008:QOS:1452520.1452548} observed that longitudinal measurement of DNSSEC deployment were critical to understand its health, and 
proposed a set of three metrics
to summarize the status of the global deployment.
Subsequent large-scale measurements~\cite{182950} reported to confirm earlier findings, but also began to raise concerns about validating resolvers and the limited
number of users being protected by DNSSEC.

More recent work~\cite{chung2017longitudinal} has conducted large-scale longitudinal analysis of the global DNSSEC deployment.  Therein, the
authors examined two years of
DNSSEC deployment data and addressed the identification of key management errors as an area for concern.
The authors also
discussed the subject of recycled keys %
as key ``sharing'' and flagged the behavior as an error.
Recycled keys are those that were shared in separate zones or used, removed, and then re-used. 
DNSSEC zones are often presumed to create distinct keys for themselves without sharing usage with other zones, and that once a key expires and completes its operational lifetime, 
it will not be used again.  
In our paper, we treated each appearance of a recycled key as a new key, because when being re-used its operational life cycle is different.
From our corpus we identify 35,882,395 distinct {\tt DNSKEY}s,
42,908,290 distinct {\tt DNSKEY}/zone pairs, and 54,337,697 total operational lifetimes of keys.
Our measurements confirm the earlier result, and further illustrate that a non-trivial number of {\tt DNSKEY} records were shared between zones, and others may have been returned to 
service after completing an initial operational lifetime.

While recent related work signals a renewed interest in key transitions, previous literature exists that suggests augmenting the DNSSEC protocol to add explicit
semantics that indicate ongoing key transitions.  Guette~\etal~\cite{guette2005algorithm} suggested an extension to the {\tt DNSKEY} format itself to
indicate when key transitions are underway.  In subsequent work~\cite{guette2009automating}, this approach was evolved by proposing the new Resource Record~{\tt KRI}.
Interestingly, the semantics that those works suggest as being necessary are already observable in the current DNSSEC, when using the methodology we introduced in this paper. Explicitly exposing key transitions in the DNSSEC may have security implications, though.
Nawrocki~\etal~\cite{njsw-fsdat-21} show that the presence of multiple keys in the DNS gets systematically exploited in DDoS attacks.  

The DNSSEC algorithm rollovers, an aspect that is outside the scope of this work, has been studied in the past~\cite{chung2017longitudinal,muller2019rolling}.
In a recent publication, M\"{u}ller~\etal~\cite{10.1145/3419394.3423638} study the lifetime of cryptographic algorithms for DNSSEC to conclude that \textit{algorithm agility} has already been reached for DNSSEC.

Finally, the role of machine learning in assessing security aspects and detecting various attacks, which has been discussed in the related work, can also be useful for our future work specifically with respect to noncompliant transitions.
Jin, Tomoishi, and Matsuura~\cite{8935025}, for example, make use of machine learning to detect cache poisoning in DNS, specifically those caused by compromised name servers.
In the context of Web PKI, Dong, Kane, and Camp~\cite{10.1145/2975591} define a set of features to describe X.509 and apply deep neural networks to detect rogue certificates.
And finally, in general context of relational data, Lou~\etal~\cite{10.1145/3447548.3467088} proposes a method to cluster related data.

\section{Discussion}\label{sec:disc-disc}

The
global rollout of DNSSEC is 
flourishing, and is giving operators experience at scale in 
managing cryptography in a core Internet protocol.
We believe that now is the right time to investigate what has been (and can be) learned from these experiences.

\paragraph{Capturing the right security model}
Caching plays an important role in DNS(SEC) because it enables scalability and availability.
This service is controlled by \texttt{TTL} values of records.
From a security perspective~\cite{andress2014basics}, when changing {\tt DNSKEY}s, care must be taken to provision consistent {\tt DNSKEY} and {\tt RRSIG} material so that what can appear in caches remains verifiable at all times (see \autoref{sec:sec-analysis}).

The DNSSEC \emph{availability} protections are important
but distinct from its data \emph{integrity} assurances.  
Availability is entirely governed by \texttt{TTL}, whereas the integrity protection is entirely governed by {\tt DNSKEY}s, {\tt RRSIG}s, and {\tt DS}es and their timings and cryptographic life-times.
TTL-based availability protections (designed for caching) are not involved in integrity assurances.
Conversely, integrity protections actually mollify 
cache poisoning attacks, which was a central design goal of DNSSEC~\cite{kaminsky2008black}.
In such attacks, availability is not a factor because adversaries influence the presence of data in caches.  
Though DNSSEC operations involve the confluence of these two aspects of DNS(SEC).
In this work, we focused on the DNSSEC integrity protections and stress the importance of evaluating their exclusive role.
Any changes to RRs can succeed within the temporal constraints imposed by DNSSEC (see \autoref{sec:sec-analysis}), and in turn, how caches regard \texttt{TTL} values has no impact on security guarantees provided by DNSSEC.
We proposed that the necessary conditions to preserve integrity protections during a DNSSEC key transition are:
\vspace{-1mm}
\begin{enumerate}
\item Zone KSKs must be covered by valid and verifiable {\tt DS}~records to establish a chain of trust.
\item {\tt RRSIG}s covering current data (\texttt{DSNKEY} and other RRSets) must be verifiable by at least one authorized {\tt DNSKEY} at all times.
\end{enumerate}
These conditions together are sufficient to provide data integrity assurance in the DNSSEC  
during key transitions.  We used our proposed anatomy to describe how to evaluate these in
operational deployments.
We also discussed (\autoref{sec:sec-analysis}) the security implications of key transitions both in terms of the relationship of a key to its own signatures, to other keys, and to delegation signers in in case of KSK.
Our proposed temporal constraints do not only allow for a formal security analysis of transitions but can also be used to design and validate software used by authoritative nameservers to manage key transitions.

\paragraph{Sampling frequency versus TTL-level polling}
There is debate in the literature on the general topic of whether to monitor
{\tt DNSKEY} transitions at the granularity of DNS \texttt{TTL}s or signature lifetimes.
Specifically, work by Chung~\etal~\cite{chung2017longitudinal} considers \texttt{TTL}
to be the primary discriminator of change frequency. 
We argue that when conducting a key transition, changing zone contents (\texttt{DNSKEY}s or otherwise) before their definitive
\texttt{RRSIG}-based lifetimes have elapsed (\eg at the granularity of \texttt{TTL}s) exposes zones to vulnerabilities (\eg replay attacks), as discussed in \autoref{sec:sec-analysis}.
This holds because the security guarantees of DNSSEC are orthogonal to \texttt{TTL}s and caching because the TTL values cannot ensure the absence of replayed  (possibly compromised) keys into validating resolver caches.
More specifically, an adversary can replay compromised keys whose \texttt{RRSIG}s are still valid, regardless of their \texttt{TTL}.
For example, if a zone private key was compromised and the operator of the zone was to replace it with a new key (\ie remove the old key immediately or when \texttt{TTL}s have expired) a zone would still be vulnerable to replay attacks.
While this does not stop operators from performing key transitions like this, one of the objectives of our analyses was to demonstrate the relationships (and gaps) between protection and practices.

While our analyses illustrate weaknesses, our measurement corpus actually includes observations taken at half the \texttt{TTL} values of the
DNS Root zone and all TLDs whose TTLs are all two days.
We posit that one-day polling of these zones is sufficient based on the
Nyquist-Shannon sampling theorem~\cite{nyquist1928certain,shannon1949communication}.
Measuring continuous phenomena with discrete sampling can approximate those phenomena and accurately characterize them, if polling occurs at 
a frequency that is at least twice the rate of change. 
This substantiates our three conclusions.
\one Key transitions need to be measured at frequencies relative to {\tt RRSIG} lifetimes. 
\two Properly operated infrastructures (\eg the Root and TLDs) perform their operations in accordance with these analyses.
\three This work is able to compare these different timing hypotheses (\ie {\tt RRSIG} vs. 
\texttt{TTL}) using longitudinal measurements from the wild.

Finally, it should be noted that discrete measurements, regardless of the measurement frequency, are going to inevitably miss alterations that happen between subsequent measurements.
With regard to our measurements, this means that violations to temporal constraints (\autoref{sec:sec-analysis}) might be overseen as we construct a continuous model using the binding, bridging, and busting method (\autoref{sec:methodology}).

\paragraph{Key transitions are complex}
The anatomy of key transitions in the wild shows a large diversity in the different ways that
zone administrators are deploying them. 
This diversity has often not conformed to those prescriptions set down in RFCs, but is that
a problem of the zone owners (and their software) or with the standards?  
Based on the $\approx$18~million transitions that we observed in this work, a more foundational question arises:
Are operator practices exposing correctable security problems or are
their implementations displaying insights that should be ingested---similar to the Continuous Improvement paradigm~\cite{deming1982out}, which has been applied recently in 
other areas~\cite{bhuiyan2005overview}?
This question serves as an additional motivation for the behavior-based classification approach we described in Section~\ref{sec:methodology}.
We believe that 
defining an anatomy and measuring related features
can serve as a rigorous methodology,
which  may  give rise to 
 a sound feedback loop between standardization and operational practice.

\section{Conclusion and Outlook}\label{sec:disc}

In this work, we 
defined the
\emph{anatomy of key transitions} 
as ten~separate, measurable timing features based on a temporal model for DNS keys.
In addition, we identified 
related measurable aspects of key life cycle management (\eg relative age and key management errors), and defined a holistic behavior-based classification method for transitions.
We then introduced a novel methodology for converting discrete observations into continuous {\tt DNSKEY} lifetimes, named \emph{Bridging, Busting, and Binding}.
Using this
substrate, we created anatomies of key transitions from the wild and empirically evaluated this complex phenomenon for the first 15 years of DNSSEC~operation.

Our work allows us to measure how well operators have followed guidance and, 
as an example of the general utility of our methodology,
where related work~\cite{wang2014emergency} has applied.
More generally, we were able to classify the kinds of key transitions operators have been using.
We found that the majority of KSK key transitions 
and the vast majority of ZSK key transitions 
do not conform to RFC guidance.
Additionally, using our behavior-based classification, we
also observed that in some years significant rates of key transitions were rolling backward.
We conclude that 
the anatomy and methodology of this work serve as a useful platform for investigating DNSSEC key transitions
in the wild, and results are well suited to inform evolving key transition practices.

As operational guidance, we advise operators should \one~aim for transition methods that reduce dependencies on parent zones and \two use the analyses of this paper to help verify that RRsets covered by keys remain verifiable when seen from external vantage points.
Public recursive resolvers may serve as such vantage points to verify the state of cached entries.

Going forward, we intend to examine the specific implications of popular key transitions, to understand where deployed innovations can advise security standards.
To better evaluate and understand noncompliant transitions, it is also possible to make use of statistical approaches common in artificial intelligence and machine learning to cluster similar transitions~\cite{10.1145/3447548.3467088} or even to engineer and select new features beside the ten introduced in this work.
Further, we intend to investigate the applicability of using our anatomy for other large-scale object security systems, such as 
the Resource Public Key Infrastructure (RPKI)~\cite{rfc6480} and the Web PKI~\cite{wpkops}.
Finally, we intend to make our feature set and the longitudinal dataset from~\cite{secspider-web} public,
to encourage additional investigations from the community.

\paragraph{Acknowledgments}
This work was supported by the Commonwealth Cyber Initiative, an investment in the advancement of cyber R\&D, innovation
and workforce development. For more information about CCI, visit \url{cyberinitiative.org}.
Additional support in parts was provided by the German Federal Ministry of Education and Research (BMBF) within the projects \emph{Deutsches Internet-Institut} (grant no. \textit{16DII111}) and \emph{PRIMEnet}.

\label{lastbodypage}

\balance
\bibliographystyle{ieeetran}
\bibliography{paper,rfc}

\clearpage

\appendices
\setcounter{table}{0}
\renewcommand\thetable{\thesection.\arabic{table}}
\setcounter{figure}{0}
\renewcommand\thefigure{\thesection.\arabic{figure}}

\onecolumn
\section{Raw Data of Monitored Zones}\label{app:zone-count}

\begin{figure}[h]
	\center
	\includegraphics[width=0.75\columnwidth,keepaspectratio]{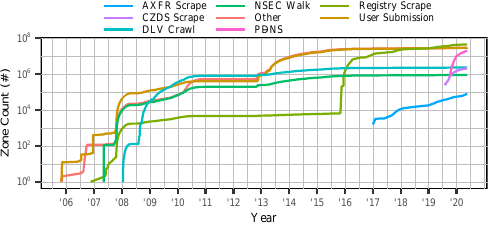}
	\caption{Total number of monitored DNSSEC-enabled zones per source and year}
	\label{fig:zone-cdf-source}
\end{figure}

The number of DNSSEC-enabled zones that we have monitord has constantly been increasing in the past 15 years.
\autoref{fig:zone-cdf-source} depicts the growth of secure zones for various sources in our data corpus.
\autoref{tab:zone-count-year-source} tabulates the absolute values per year and source.

\begin{table*}[!h]
	\setlength{\tabcolsep}{4pt}
	\scriptsize
	\caption{Total number of monitored DNSSEC-enabled zones per source.}
	\label{tab:zone-count-year-source}
	\begin{tabularx}{\textwidth}{X r r r r r r r r r r r r r r r r}
		\toprule
		& \multicolumn{16}{c}{Year}
		\\
		\cmidrule{2-17}
		& \makecell[c]{'06} & \makecell[c]{'07} & \makecell[c]{'08} & \makecell[c]{'09} & \makecell[c]{'10} & \makecell[c]{'11} & \makecell[c]{'12} & \makecell[c]{'13} & \makecell[c]{'14} & \makecell[c]{'15} & \makecell[c]{'16} & \makecell[c]{'17} & \makecell[c]{'18} & \makecell[c]{'19} & \makecell[c]{'20}\\
		\midrule
		\input{./stats/zone-per-source-year.dat}
		\bottomrule
	\end{tabularx}
\end{table*}

\section{Raw Data of Key Transition Classifications}\label{app:corpus}

Throughout the paper we normalize our data to better distinguish and highlight trends.
Tables~\ref{tab:raw-key-errors}--\ref{tab:raw-key-behaviour-error} tabulate the raw numbers used to generate Figures~\ref{fig:yearly-err-perc}--\ref{fig:trans-ssclass-per-year}.

\begin{table*}[h]
	\setlength{\tabcolsep}{4pt}
	\scriptsize
	\caption{Total number of key management errors.}
	\label{tab:raw-key-errors}
	\begin{tabularx}{\textwidth}{X r r r r r r r r r r r r r r r}
		\toprule
		& \multicolumn{15}{c}{Year}
		\\
		\cmidrule{2-16}
		& \makecell[c]{'06} & \makecell[c]{'07} & \makecell[c]{'08} & \makecell[c]{'09} & \makecell[c]{'10} & \makecell[c]{'11} & \makecell[c]{'12} & \makecell[c]{'13} & \makecell[c]{'14} & \makecell[c]{'15} & \makecell[c]{'16} & \makecell[c]{'17} & \makecell[c]{'18} & \makecell[c]{'19} & \makecell[c]{'20}\\
		\midrule
		\input{./stats/key-errors-per-year-barplot-raw-body.dat}
		\bottomrule
	\end{tabularx}
\end{table*}

\begin{table*}[h]
	\setlength{\tabcolsep}{4pt}
	\scriptsize
	\caption{RFC-based classification: Number of different key transitions for KSK (top) and ZSK (bottom).}
	\label{tab:raw-key-rfc-class}
	\begin{tabularx}{\textwidth}{X r r r r r r r r r r r r r r r}
		\toprule
		& \multicolumn{15}{c}{Year}
		\\
		\cmidrule{2-16}
		& \makecell[c]{'06} & \makecell[c]{'07} & \makecell[c]{'08} & \makecell[c]{'09} & \makecell[c]{'10} & \makecell[c]{'11} & \makecell[c]{'12} & \makecell[c]{'13} & \makecell[c]{'14} & \makecell[c]{'15} & \makecell[c]{'16} & \makecell[c]{'17} & \makecell[c]{'18} & \makecell[c]{'19} & \makecell[c]{'20}\\
		\midrule
		\input{./stats/ksk-rfcclass-per-year-barplot-raw-body.dat}
		\midrule
		\input{./stats/zsk-rfcclass-per-year-barplot-raw-body.dat}
		\bottomrule
	\end{tabularx}
\end{table*}

\begin{table*}[h]
	\setlength{\tabcolsep}{4pt}
	\scriptsize
	\caption{Number of valids, warnings, and errors for KSK (top) and ZSK (bottom; excluding type noncompliant).}
	\label{tab:raw-key-trans-error}
	\begin{tabularx}{\textwidth}{X r r r r r r r r r r r r r r r}
		\toprule
		& \multicolumn{15}{c}{Year}
		\\
		\cmidrule{2-16}
		& \makecell[c]{'06} & \makecell[c]{'07} & \makecell[c]{'08} & \makecell[c]{'09} & \makecell[c]{'10} & \makecell[c]{'11} & \makecell[c]{'12} & \makecell[c]{'13} & \makecell[c]{'14} & \makecell[c]{'15} & \makecell[c]{'16} & \makecell[c]{'17} & \makecell[c]{'18} & \makecell[c]{'19} & \makecell[c]{'20}\\
		\midrule
		\input{./stats/ksk-trans-err-per-year-barplot-raw-body.dat}
		\midrule
		\input{./stats/zsk-trans-err-per-year-barplot-raw-body.dat}
		\bottomrule
	\end{tabularx}
\end{table*}

\begin{table*}[h]
	\setlength{\tabcolsep}{4pt}
	\scriptsize
	\caption{Behavior-based classification: Number of different key transitions for KSK (top) and ZSK (bottom).}
	\label{tab:raw-key-behaviour-error}
	\begin{tabularx}{\textwidth}{X r r r r r r r r r r r r r r r}
		\toprule
		& \multicolumn{15}{c}{Year}
		\\
		\cmidrule{2-16}
		& \makecell[c]{'06} & \makecell[c]{'07} & \makecell[c]{'08} & \makecell[c]{'09} & \makecell[c]{'10} & \makecell[c]{'11} & \makecell[c]{'12} & \makecell[c]{'13} & \makecell[c]{'14} & \makecell[c]{'15} & \makecell[c]{'16} & \makecell[c]{'17} & \makecell[c]{'18} & \makecell[c]{'19} & \makecell[c]{'20}\\
		\midrule
		\input{./stats/ksk-ssclass-per-year-barplot-raw-body.dat}
		\midrule
		\input{./stats/zsk-ssclass-per-year-barplot-raw-body.dat}
		\bottomrule
	\end{tabularx}
\end{table*}

\clearpage
\section{Transitions in the Wild}\label{app:wild}

\begin{figure*}[h]
\begin{center}
    \begin{subfigure}[t]{0.8\textwidth}
      \begin{center}
        \includegraphics[width=\textwidth]{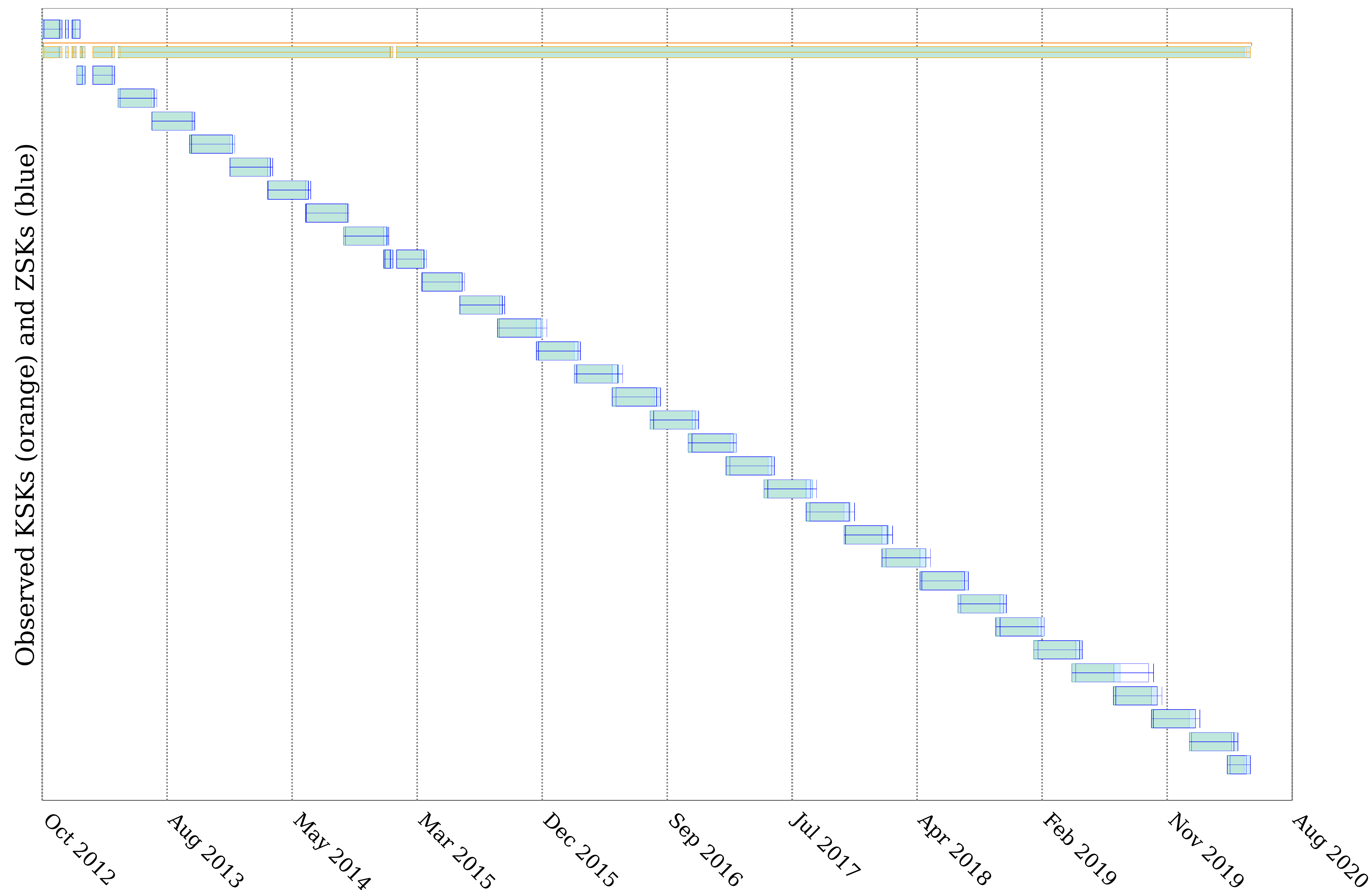}
        \caption{The key management of the {\tt .com} zone shows exemplary regularity and structure in its key transitions.}
        \label{fig:com-only-trans-com}
      \end{center}
    \end{subfigure}
    \hfil
    \begin{subfigure}[t]{0.8\textwidth}
      \begin{center}
        \includegraphics[width=\textwidth]{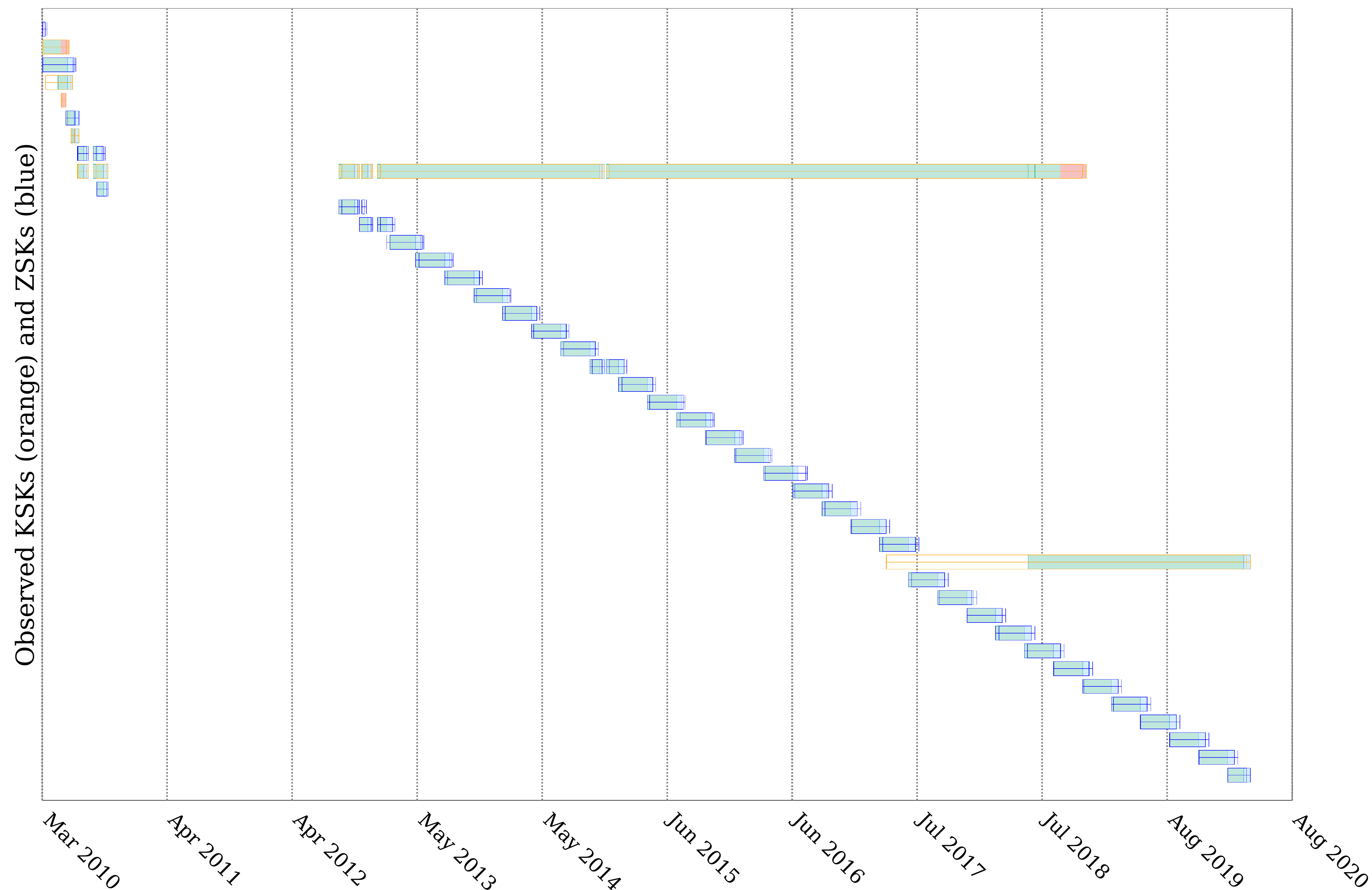}
        \caption{Shown here is the Root zone's KSK transition away from the DURZ key (followed by revocation), and regular periodicity in its ZSK transitions.}
        \label{fig:root-only-trans-root}
      \end{center}
    \end{subfigure}
    \caption{The key transition behaviors of two prominent DNSSEC zones.}
    \label{fig:wild-prominent-zones}
\end{center}
\end{figure*}

\begin{figure*}
\begin{center}
    \begin{subfigure}[t]{0.45\textwidth}
      \begin{center}
        \includegraphics[width=\textwidth]{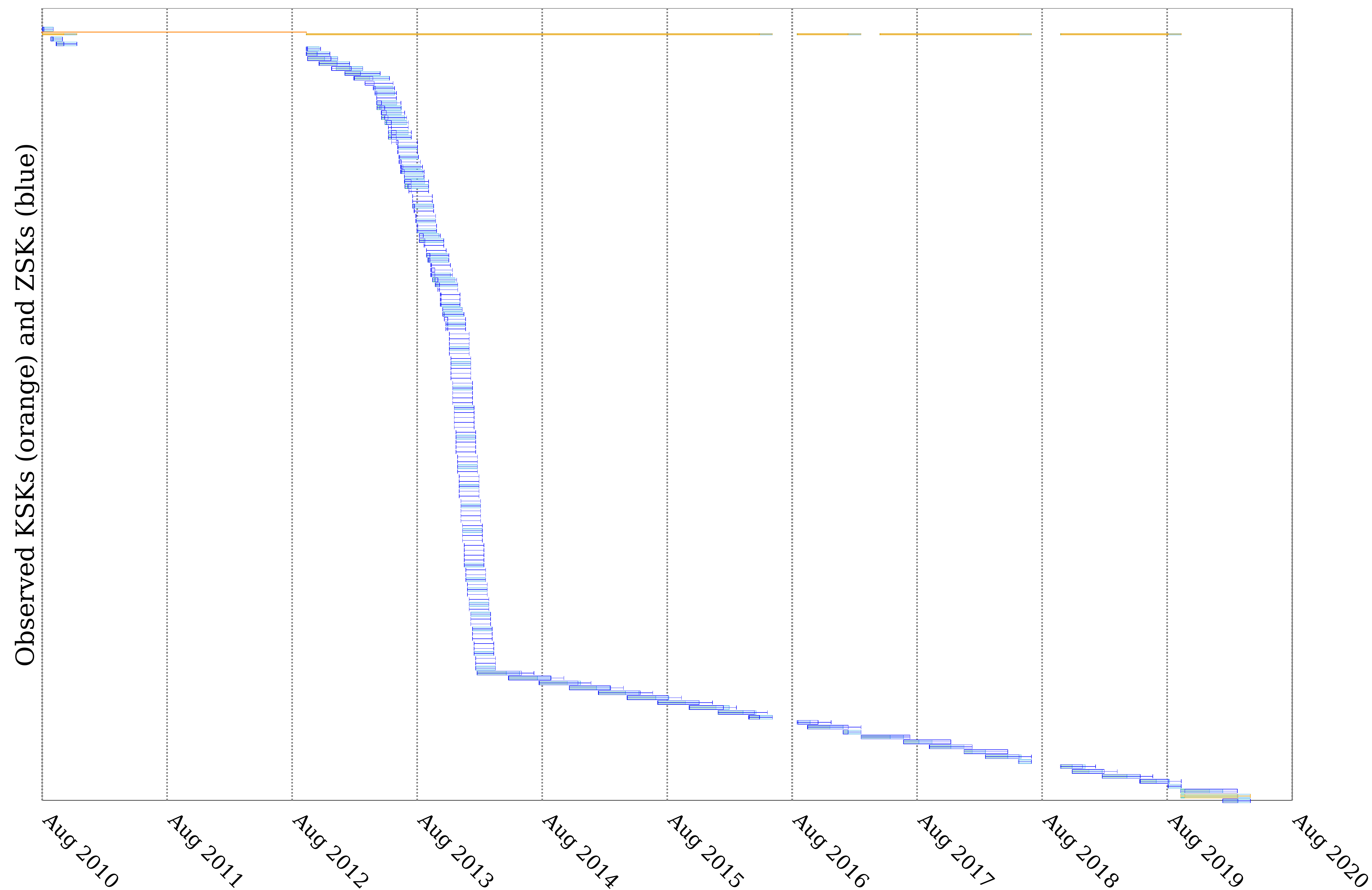}
        \caption{{\tt softwarehotels.se} shows distinctive key transition behavior, with up to \underline{55} valid keys (though not all in use at the same time).}
        \label{fig:softwarehotels.se-trans}
      \end{center}
    \end{subfigure}
    \hfil
    \begin{subfigure}[t]{0.45\textwidth}
      \begin{center}
        \includegraphics[width=\textwidth]{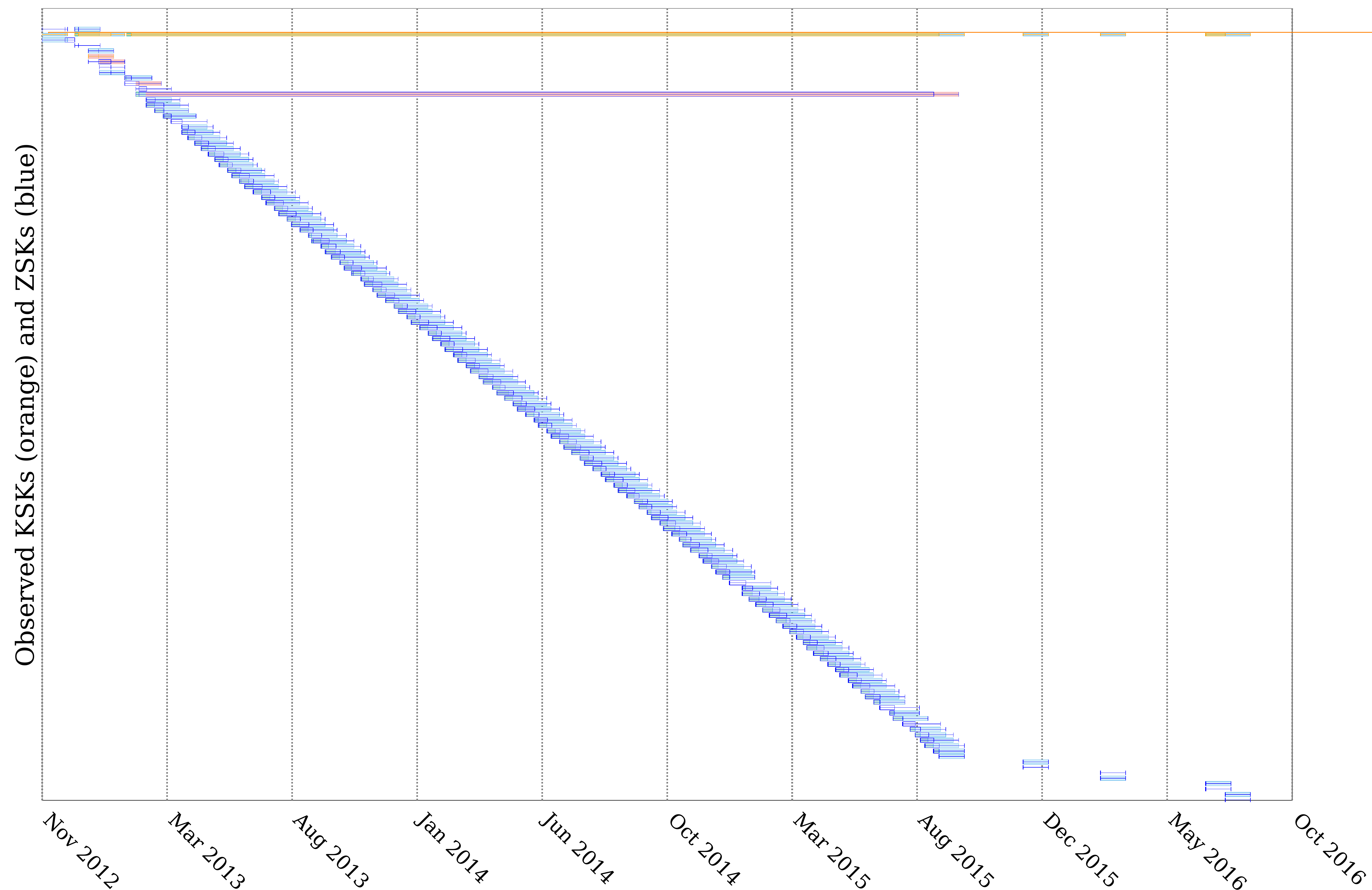}
        \caption{The zone {\tt cyrrax.com} shows high-frequency transitions of its ZSKs, with regular periodicity and extended \emph{Retire} times.}
        \label{fig:cyrrax.com-trans}
      \end{center}
    \end{subfigure}
    \\
    \begin{subfigure}[t]{0.45\textwidth}
      \begin{center}
        \includegraphics[width=\textwidth]{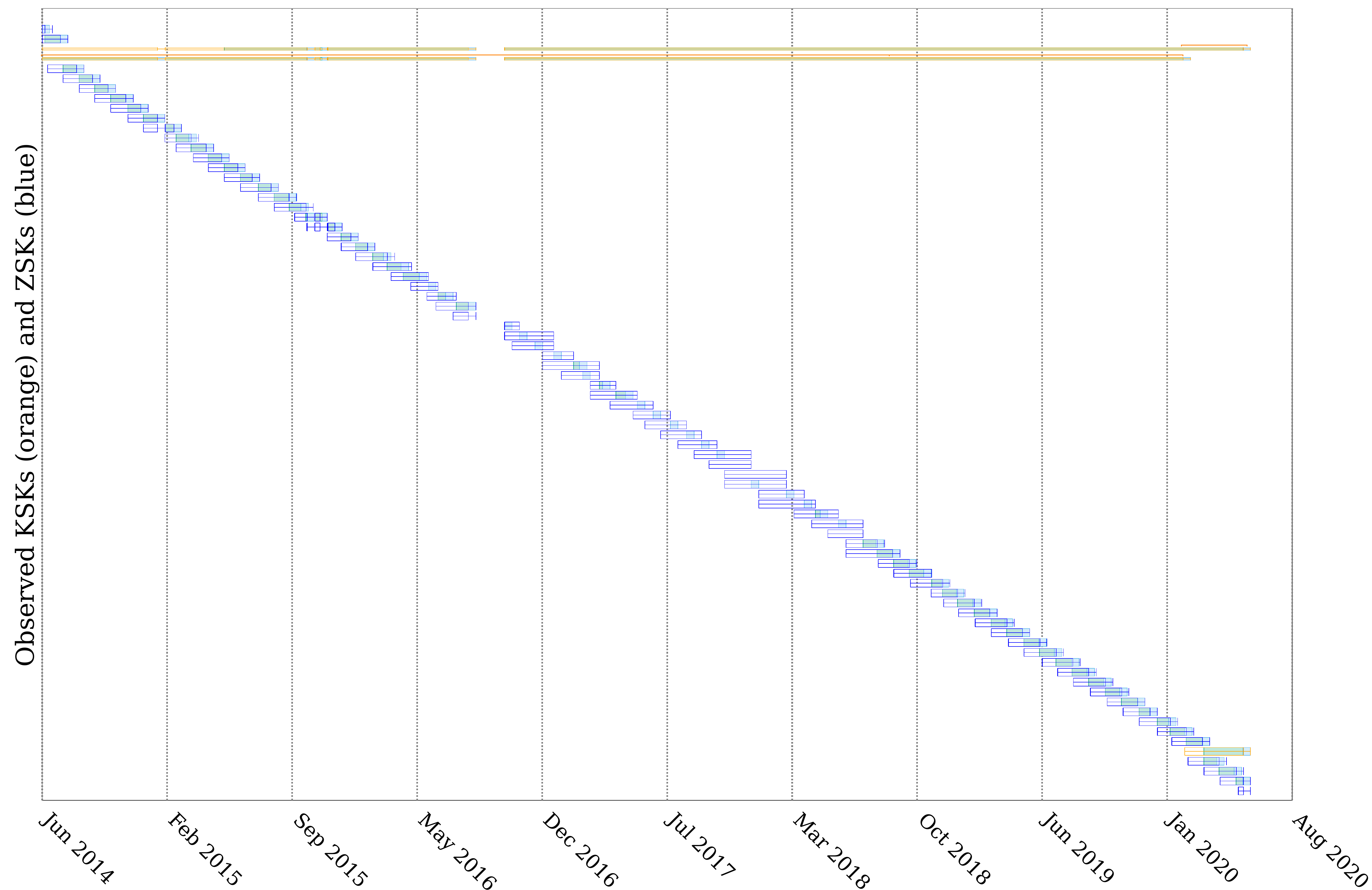}
        \caption{The key management of ARIN (a Regional Internet Registry, RIR), uses extended \emph{PreStage} periods for its keys (ZSKs and KSKs).}
        \label{fig:arin-net-only-trans-keyroll}
      \end{center}
    \end{subfigure}
    \hfil
    \begin{subfigure}[t]{0.45\textwidth}
      \begin{center}
        \includegraphics[width=\textwidth]{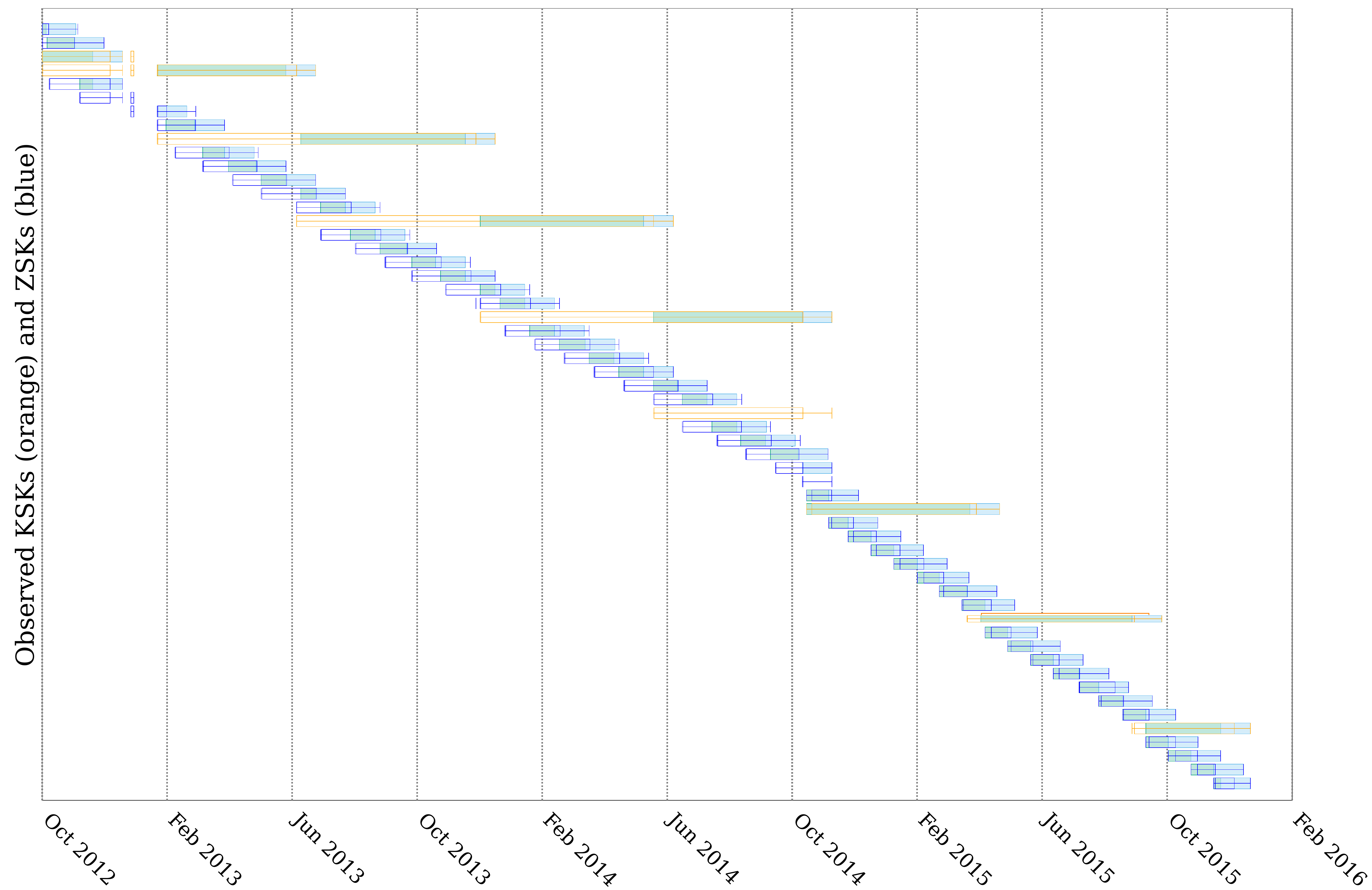}
        \caption{The key management of {\tt advantage.gov} visibly changes from having protracted \emph{PreStage} periods to a different model at the end of 2014.}
        \label{fig:advantage.gov-trans}
      \end{center}
    \end{subfigure}
	\begin{subfigure}[t]{0.45\textwidth}
		\begin{center}
			\includegraphics[width=\textwidth]{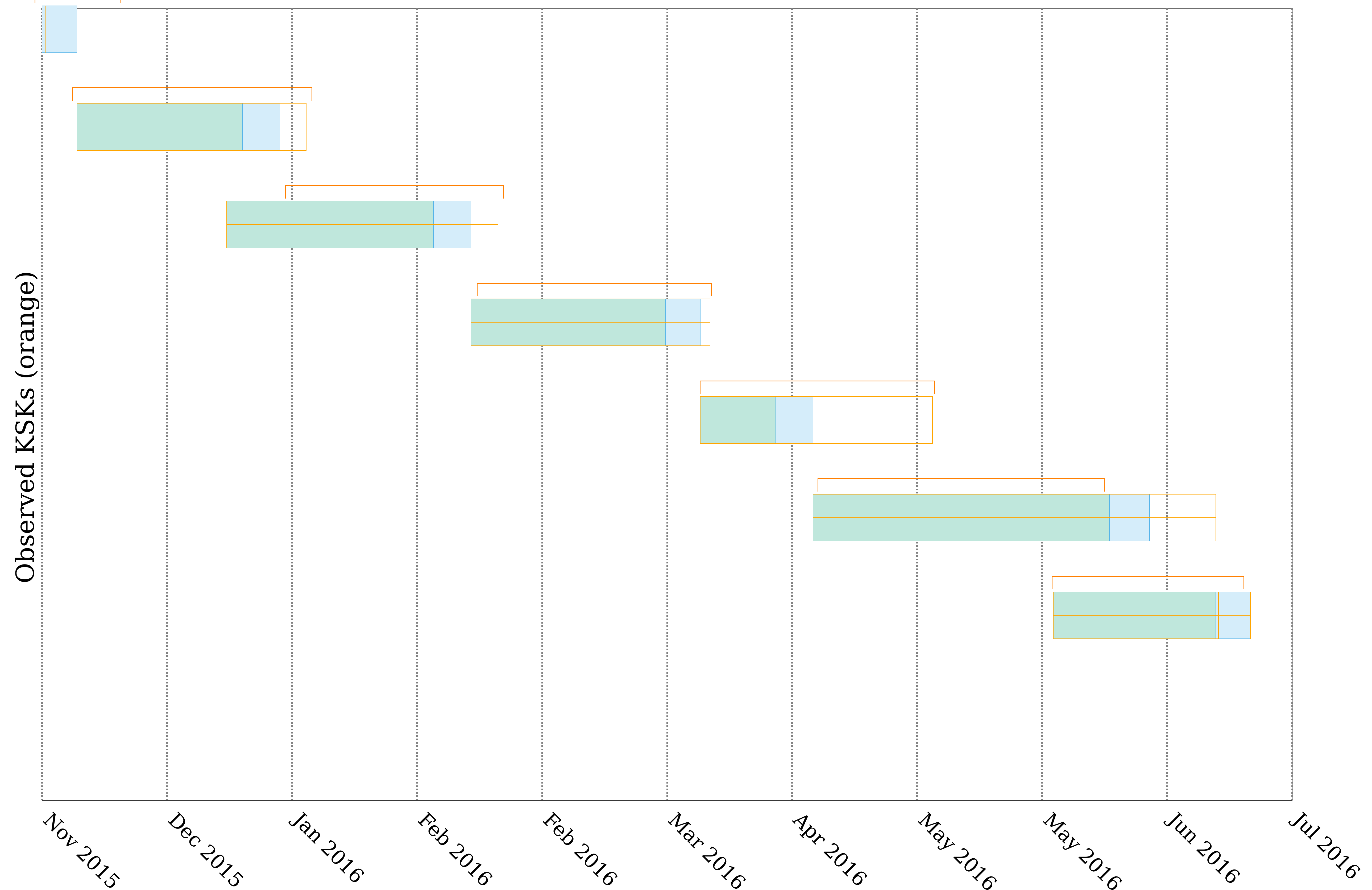}
			\caption{The key management of {\tt willemvanveldhuizen.com} exhibits regular DoubleDS transitions without having any ZSKs in the zone.}
			\label{fig:willemvanveldhuizen.com-trans}
		\end{center}
	\end{subfigure}
    \caption{Key transition patterns from other operational zones in the wild.}
    \label{fig:wild-zones}
\end{center}
\end{figure*}

To illustrate the complete key life cycles of zones, Figures~\ref{fig:wild-prominent-zones} and~\ref{fig:wild-zones} include both ZSKs (in blue) and KSKs (in yellow).
Red sections indicate periods in which keys have the revoke bit set~\cite{rfc5011}.  
Notable in these (and other) zones is the difference of life cycles of ZSKs and KSKs.

\twocolumn

\clearpage
\nobalance
\section*{Author Biography}

\begin{IEEEbiography}[{\includegraphics[width=1in,height=1.25in,clip,keepaspectratio]{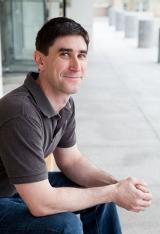}}]{Eric Osterweil}
  is an assistant professor of Computer Science at George Mason University, where he directs the Measurable Security Lab (MSL).
  He received his Ph.D. from the University of California, Los Angeles (UCLA)
  in 2010.  After graduating, he joined the Verisign Labs research team where he became a principal scientist working in the Office of the Chief
  Security Officer (CSO).  His research foci are the cybersecurity of Internet critical infrastructures, inter-organizational object-security,
  and Internet measurements.
  He is the former co-chair/vice-chair of ICANN's Second Security, Stability, and Resiliency Review Team (SSR2 RT), an active member of the IETF,
  the maintainer of multiple open source software projects (\eg the DANE reference library libCanute, the high-speed DNS polling library libVantages,
  and more), and is the inceptor (and ongoing curator) of the oldest dataset of DNSSEC measurement dataset (SecSpider).
\end{IEEEbiography}

\begin{IEEEbiography}[{\includegraphics[width=1in,height=1.25in,clip,keepaspectratio]{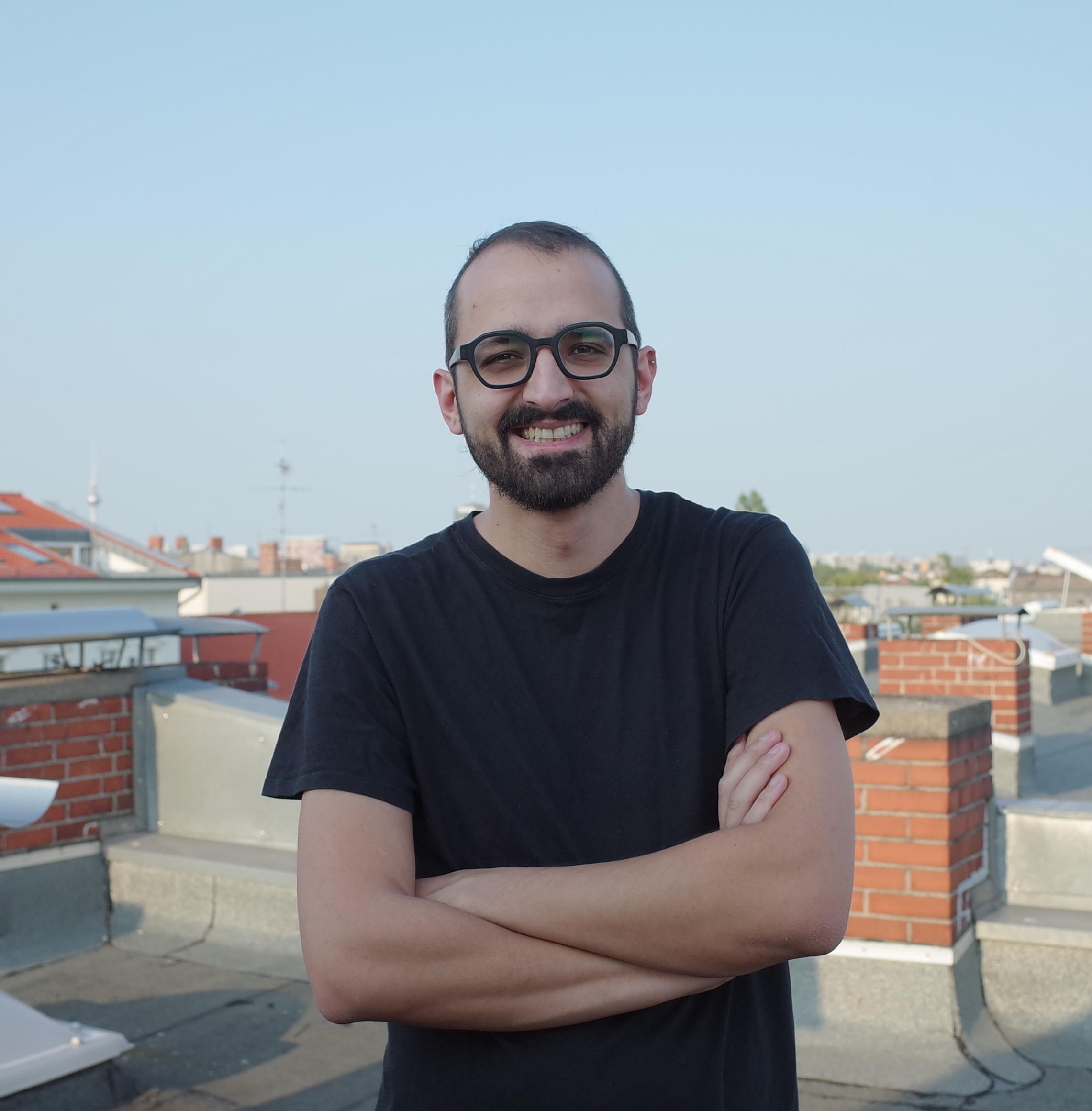}}]{Pouyan Fotouhi Tehrani}
is a Ph.D. candidate of Computer Science at The Weizenbaum Institute for the Networked Society under the supervision of Matthias W\"ahlisch and Thomas C. Schmidt.
His research evolves around communication networks for emergency and crisis with a focus on object security in fragmented and challenged networks.
With publications on namespace management in Information Centric Networks (ICN), X.509-based identification on the Web, and security of DNSSEC, his research addresses both technical and organizational aspects of existing security approaches on the Internet and their proper adaptation for the next generation Internet, \eg named-data networks.
\end{IEEEbiography}

\vfill

\newpage

\begin{IEEEbiography}[{\includegraphics[width=1in,height=1.25in,clip,keepaspectratio]{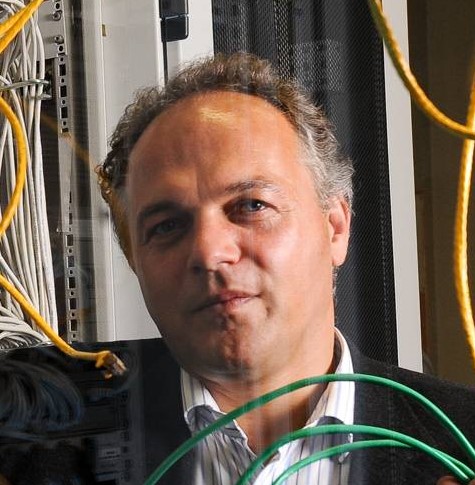}}]{Thomas C. Schmidt}
is professor of Computer Networks and Internet Technologies
at Hamburg University of Applied Sciences (HAW), where he heads the Internet
Technologies research group (iNET). Prior to moving to Hamburg, he was director of a scientific computer centre in Berlin. He studied mathematics, physics and German literature at Freie Universitaet Berlin and University of Maryland, and received his Ph.D. from FU Berlin in 1993. Since then he has continuously conducted numerous national and international research projects. He was the principal investigator in a number of EU, nationally funded and industrial projects as well as visiting professor at the University of Reading, U.K.. His continued interests lie in the development, measurement, and analysis of large-scale distributed systems like the Internet. He serves as co-editor and technical expert in many occasions and is actively involved in the work of IETF and IRTF. Together with his group he pioneered work on an information-centric Industrial IoT and the emerging data-centric Web of Things. Thomas is a co-founder of several large open source projects and coordinator of the community developing the RIOT operating system - the friendly OS for the Internet of Things.
\end{IEEEbiography}

\begin{IEEEbiography}[{\includegraphics[width=1in,height=1.25in,clip,keepaspectratio]{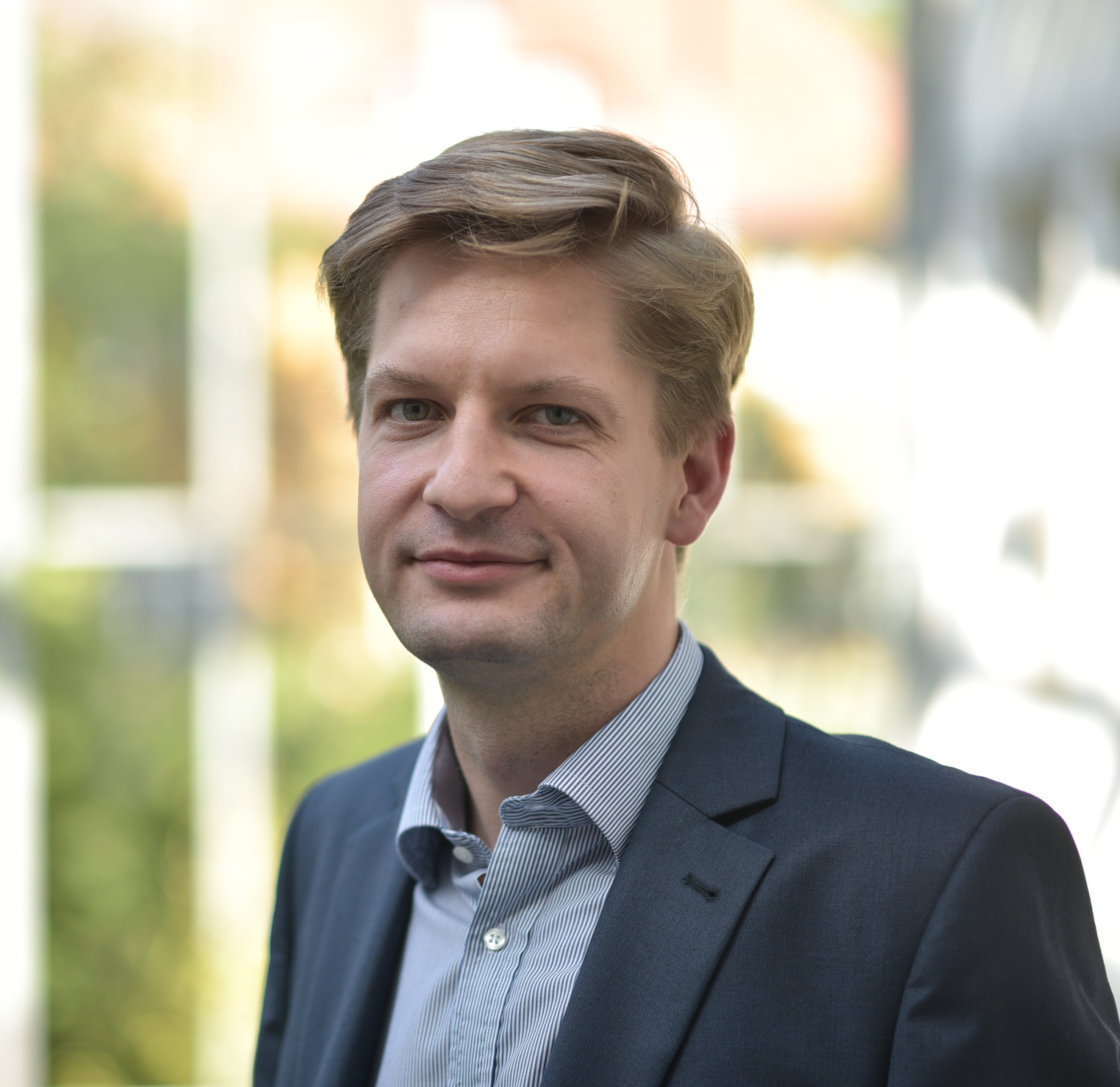}}]{Matthias W{\"a}hlisch}
is an assistant professor of Computer Science at Freie Universit{\"a}t Berlin, heading the Internet Technologies Research group.
He received his Ph.D. in computer science with highest honors from Freie Universit{\"a}t Berlin.
His research and teaching focus on efficient, reliable, and secure Internet communication.
This includes the design and evaluation of networking protocols and architectures as well as Internet measurements and analysis.
His efforts are driven by the goal to improve Internet communication based on sound research.
Matthias is the PI of several national and international projects, supported by overall 5.1M EUR grant money.
He published more than 150 peer-reviewed papers (\eg at ACM HotNets, ACM CoNEXT, ACM IMC, USENIX Security, The Web Conference).
 Since 2005, Matthias is actively involved in Internet standardization (IETF/IRTF).
His research results have been distinguished multiple times.
Amongst others, he received the Young Talents Award of Leibniz-Kolleg Potsdam for outstanding achievements in advancing the Internet, as well as the Excellent Young Scientists Award (10,000 EUR) for his contributions to the Internet of Things and their prospective entrepreneurial practice.
He received the Best of ACM SIGCOMM CCR Award in 2018 and 2019.
He co-founded and still coordinates some successful open source projects in the context of Internet of Things (RIOT) and secure Internet routing (\eg RTRlib).
\end{IEEEbiography}

\vfill

\label{lastpage}

\end{document}